\documentclass{article}

\usepackage{microtype}
\usepackage{graphicx}
\usepackage{subcaption}
\usepackage{booktabs} 

\usepackage{hyperref}

\usepackage[preprint]{icml2026}

\usepackage{amsmath}
\usepackage{amssymb}
\usepackage{mathtools}
\usepackage{amsthm}
\usepackage{multirow} 
\usepackage{pifont}   
\usepackage{colortbl}
\usepackage[capitalize,noabbrev]{cleveref}

\theoremstyle{plain}

\theoremstyle{definition}

\theoremstyle{remark}

\usepackage[textsize=tiny]{todonotes}

\icmltitlerunning{PAL: Audio Information Transfer into LLMs}

\begin{document}


\twocolumn[
  \icmltitle{PAL: Probing Audio Encoders via LLMs \\
    - Audio Information Transfer into LLMs}
    


  \icmlsetsymbol{equal}{*}

\begin{icmlauthorlist}
\icmlauthor{Tony Alex}{equal,cvssp,pai}
\icmlauthor{Wish Suharitdamrong}{pai}
\icmlauthor{Sara Ahmed}{cvssp,pai}
\icmlauthor{Armin Mustafa}{cvssp,pai}
\icmlauthor{Philip J. B. Jackson}{cvssp,pai}
\icmlauthor{Imran Razzak}{mbzuai}
\icmlauthor{Muhammad Awais}{cvssp,pai}
\end{icmlauthorlist}

\icmlaffiliation{cvssp}{Centre for Vision, Speech and Signal Processing (CVSSP), University of Surrey, UK}
\icmlaffiliation{pai}{Surrey Institute for People-Centred AI, University of Surrey, Guildford, GU2 7XH, UK}
\icmlaffiliation{mbzuai}{Mohamed bin Zayed University of Artificial Intelligence (MBZUAI), Abu Dhabi, UAE}

\icmlcorrespondingauthor{Tony Alex}{t.alex@surrey.ac.uk}

  \icmlkeywords{Machine Learning, ICML}

  \vskip 0.3in
]



\printAffiliationsAndNotice{}  

\begin{abstract}
Integration of audio perception into large language models (LLMs) is an emerging research area for enabling machine-listening applications, yet efficient transfer of rich audio semantics from audio encoders to LLMs remains underexplored. The most widely used integration paradigm projects audio-encoder output tokens into the LLM input space (e.g., via an MLP or a Q-Former) and then \emph{prepends or inserts} them into the text token sequence. We refer to this generic scheme as \emph{Prepend to the LLM's input token space (PLITS)} integration. We propose an efficient alternative, \textbf{L}ightweight \textbf{A}udio \textbf{LLM} Integration (\textbf{LAL}). LAL injects audio representations solely through the attention mechanism at selected LLM layers, bypassing the feed-forward module. It encodes rich audio semantics at an appropriate level of abstraction for integration into different transformer blocks, substantially reducing computational overhead compared to existing approaches. We further introduce \textbf{PAL}, a hybrid integration approach for efficiently \textbf{P}robing \textbf{A}udio encoders via \textbf{LLM}. \textbf{PAL} applies PLITS only to a compact set of summary tokens while integrating the full audio token sequence via LAL. Under an identical training curriculum, \textbf{LAL} consistently matches or outperforms existing integration approaches across multiple base LLMs and tasks, with improvements of up to 30\% over a strong PLITS baseline, while reducing memory usage by about 60\% and increasing throughput by about 190\%. Moreover, \textbf{PAL} matches or exceeds PLITS performance while offering substantially better computational and memory efficiency.

\end{abstract}

\section{Introduction}

Large Language Models (LLMs)~\citep{brown2020language,grattafiori2024llama,jiang2024mixtral,liu2024deepseek} have emerged as the foundational technology for natural language interaction with machines, demonstrating remarkable conversational fluency. Despite this success, their perceptual capabilities remain limited primarily to text, restricting their ability to understand the physical world. This limitation has inspired significant research into multi-modal LLMs (MLLMs), which expand traditional LLMs by integrating additional sensory modalities such as vision (Vision LLMs)~\citep{liu2023visual,templeton2024scaling,wang2024cogvlm}, audio (Large Audio Language Models (LALMs) or simply audio-LLMs)~\citep{deshmukh2023pengi,gong2024listen,tang2024salmonn,ghosh2024gama,ghosh2025audio}, and other inputs~\citep{brohan2023rt,thawkar2023xraygpt} to foster more natural, intuitive, and effective human-machine interfaces.

An audio LLM typically comprises three components: (i) a large language model (LLM), (ii) audio encoder that convert raw audio to audio features or tokens, and (iii) a mechanism that integrates encoder outputs into the LLM. 
When it comes to the integration of audio encoders with the LLM, two architectural paradigms dominate today. The first transforms the outputs of an audio encoder or encoders into the LLM input space (e.g., via an MLP, a QFormer~\citep{li2023blip}, etc.), then \emph{prepend or insert} these audio tokens to the text tokens and propagates the entire sequence through all LLM layers as if decoding jointly over audio and text. Please note that the common theme in this family is how audio tokens are passed to the LLM: they are \emph{prepended} to the text tokens. We refer to this generic scheme \textbf{Prepend to the LLM’s input token space (PLITS)} integration, a term we have introduced to group many state of the art methods in this family of audio LLMs such as ~\citet{wu2025step,xu2025qwen25omnitechnicalreport,chu2024qwen2audiotechnicalreport,goel2025audio,chu2023qwenaudioadvancinguniversalaudio,ghosh2024gama,tang2024salmonn,gong2024listen,deshmukh2023pengi}. The second paradigm, \textbf{Flamingo style} architectures~\citep{alayrac2022flamingo,kong2024audio}, instead insert cross attention and feedforward (FFN) blocks \emph{between} successive LLM layers; at each insertion, text tokens attend to a set of latent audio tokens, pass through the block FFN, and only then proceed to the next LLM layer. While this design improves attention efficiency relative to PLITS concatenation, the interleaved cross attention plus FFN stacks increase sequential depth and per layer compute, which can slow the forward pass.

In contrast, we introduce \textbf{LAL}, a lightweight integration that injects audio tokens into the LLM’s attention blocks \emph{as keys and values only} (without forming audio queries) and \emph{bypasses the LLM FFNs for audio tokens}. 
This reduces the attention complexity from $\mathcal{O}\!\big((N_a{+}N_t)^2\big)$ to $\mathcal{O}\!\big((N_a{+}N_t)N_t\big)$, 
where $N_a$ and $N_t$ denote the numbers of audio and text tokens, respectively. 
Since typically $N_a \gg N_t$, this yields substantial efficiency gains. 
By avoiding both quadratic attention over audio tokens and their passage through LLM FFNs, LAL substantially reduces memory usage and computation.
Unlike parameter-efficient methods such as LoRA, this is a core architectural modification, so the efficiency benefits are realized not only during training but also at inference time.

\begin{figure*}[t] 
\centering 

\includegraphics[width=0.75\textwidth]{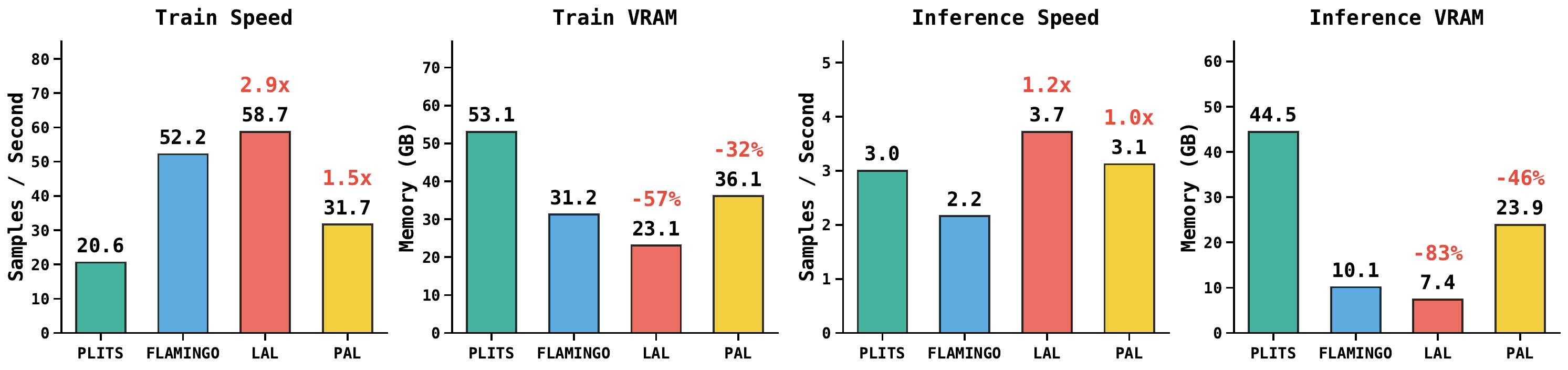}

\caption{Comparison of compute efficiency between \textbf{PLITS, state of the art audio-LLM integration} (our baseline), ~\textbf{Flamingo}, ~\textbf{LAL}(ours) and ~\textbf{PAL}(Ours). Training was performed with batch size 8 on an NVIDIA A100 using bfloat16, and inference with batch size 12 on an NVIDIA A100 using float16. All benchmarks were executed sequentially on the same node to eliminate load-related discrepancies.}
\label{fig_pal_comp}
\end{figure*}

LAL provides a compute- and memory-efficient mechanism by constraining how audio tokens interact with the LLM, while achieving performance comparable to PLITS. We further introduce a hybrid variant, \textbf{PAL}, which combines LAL and PLITS to balance efficiency and performance. \textbf{PAL} outperforms PLITS while reducing computational and memory requirements.

To validate these architectural choices, we conduct a systematic empirical study under a standardized training curriculum and dataset setup, ensuring fair comparisons across models. Our experiments explore the trade-off between performance and efficiency, highlighting how different integration techniques facilitate effective information transfer from audio encoders to LLMs with minimal parameter overhead. This analysis provides actionable insights into the design of scalable and efficient audio LLMs.

\textbf{Our main contributions are as follows: (1)} We introduce \textbf{LAL}, a lightweight integration strategy for audio-LLMs that incorporates audio tokens solely as keys and values in the LLM’s attention sub-modules and skips FFNs, thereby reducing computation and memory cost while retaining performance comparable to PLITS integration,    
\textbf{(2)} We propose \textbf{PAL}, a hybrid integration that applies PLITS to a compact set of summary tokens and LAL to the full audio token sequence, significantly outperforming PLITS while remaining efficient, and
\textbf{(3)} We conduct \textbf{fair and rigorous architectural comparisons} under a standardized training curriculum and dataset setup, providing actionable insights into the efficiency–performance trade-offs of audio-LLM design.

\section{Literature review}
\label{sec_lit_review_audio_llm_arch}

\textbf{Audio LLM architectures:} When integrating audio encoders with an LLM, two paradigms dominate. In PLITS, encoder features are mapped to the LLM token space with a small projector such as an MLP or a Q Former, the resulting audio tokens are typically prepended to the text tokens, and the joint sequence is processed by all LLM layers~\citep{wu2025step,xu2025qwen25omnitechnicalreport,chu2024qwen2audiotechnicalreport,goel2025audio,chu2023qwenaudioadvancinguniversalaudio,ghosh2024gama,tang2024salmonn,gong2024listen,deshmukh2023pengi}. In contrast, the Flamingo style architecture inserts cross attention and feed forward adapters between successive LLM layers so that text tokens attend to latent audio tokens at selected depths~\citep{alayrac2022flamingo,kong2024audio}. This makes audio to text interaction explicit and gated, but adds sequential depth, per layer compute, and parameters. An extended literature review is provided in Appendix~\ref{apx_ext_lit_review}.

\begin{figure*}[t] 
\centering 
\includegraphics[width=0.70\textwidth]{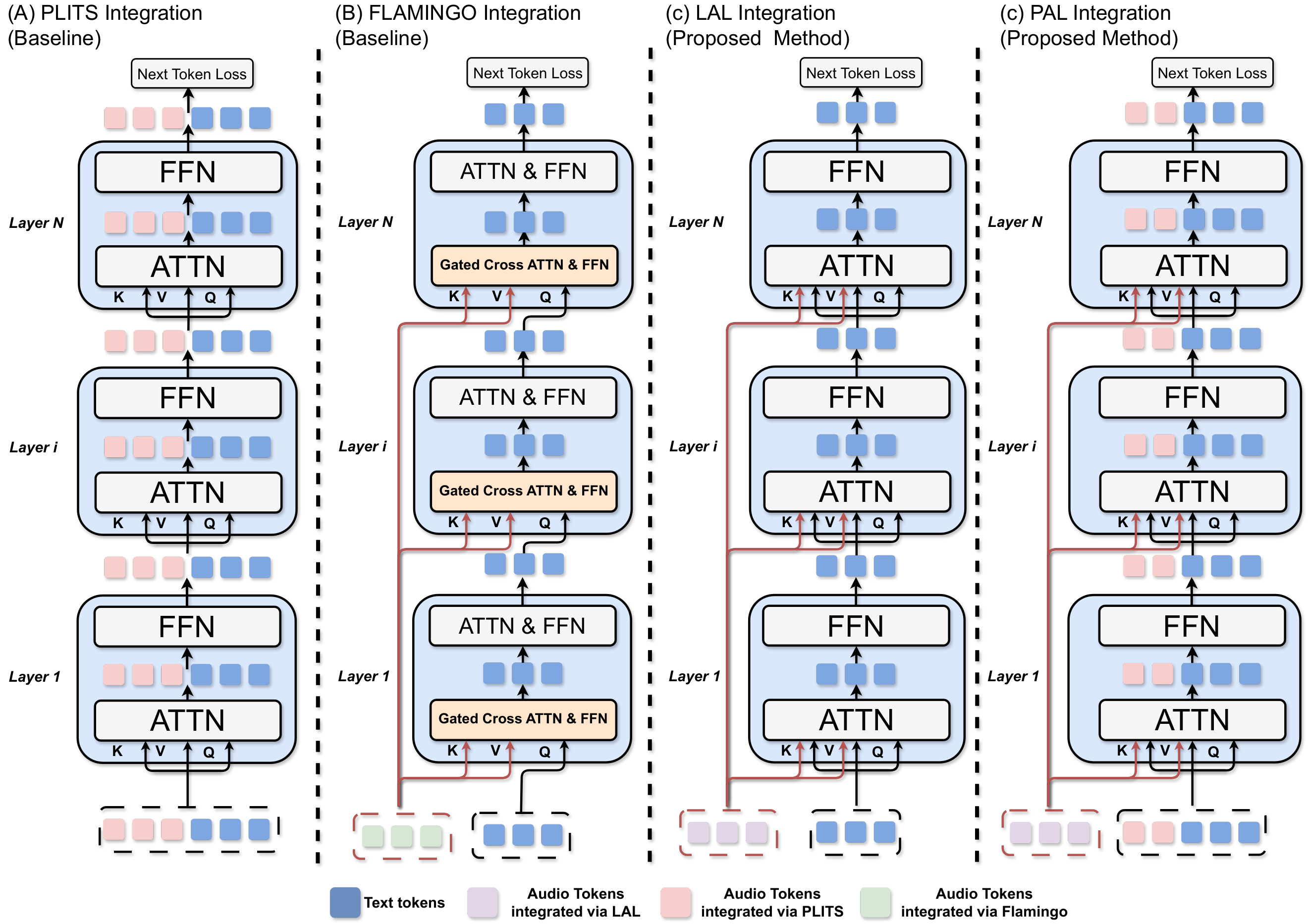} 

\caption{Illustration of integration techniques:
(A) SOTA integration \textbf{PLITS} (prepend to the LLM’s input token space), which prepends audio tokens to text tokens and propagates the full sequence through all LLM layers (our baseline);
(B) ~\textbf{Flamingo} integration, where text tokens first attend to audio tokens through a separate cross attention plus FFN module, and the resulting signal is added to the text residual stream before the next LLM layer.
(C) our proposed lightweight integration \textbf{LAL}, which introduces audio representations only through the attention mechanism (see Equations~\ref{eq_audio_insert}, \ref{eq_qkv_proj}, and \ref{eq_attn}) while bypassing the feedforward modules;
(D) \textbf{PAL}, a hybrid integration that combines \textbf{LAL} and \textbf{PLITS} integrations.}
\label{fig_pal_arch}
\end{figure*}

\section{Methodology}
This section outlines our approach to integrating audio with language models. We begin by formalizing \textbf{PLITS}, the SOTA audio-LLM integration, as our reference baseline. We then introduce \textbf{LAL}, a lightweight alternative that injects audio through attention only, and we analyze its compute and memory profile. 
Finally, we connect these findings to \textbf{PAL}, a hybrid integration that applies PLITS to a compact set of summary tokens and LAL to the full audio token sequence, significantly outperforming PLITS while remaining efficient.

\subsection{Baseline Audio LLM: Prepend to the LLM's input token space (PLITS)}
To provide a fair comparison point for our integration methods, we construct a baseline audio LLM that follows the widely adopted SOTA integration strategy, which we refer to as \emph{Prepend to the LLM's input token space (PLITS)}. In this design, the audio encoder outputs are first mapped into the LLM input embedding space using a Q-Former–style connector. The resulting audio tokens are then \emph{prepended} to the text tokens, and the concatenated sequence is passed through all LLM layers so that decoding proceeds jointly over audio and text (see Fig.~\ref{fig_pal_arch}(A)). 

The central characteristic of this PLITS-style integration is that \textbf{the audio tokens are \emph{prepended} to the text tokens}. This integration strategy is used by most audio LLMs, including several state of the art systems~\citet{wu2025step,xu2025qwen25omnitechnicalreport,chu2024qwen2audiotechnicalreport,goel2025audio,chu2023qwenaudioadvancinguniversalaudio,ghosh2024gama,tang2024salmonn,gong2024listen,deshmukh2023pengi}.

\subsection{LAL: Lightweight Audio-LLM Integration}
\label{sec_lal}
Recent work in mechanistic interpretability suggests that LLMs encode semantics as features that can be selectively activated within hidden states~\citep{elhage2022superposition,bricken2023monosemanticity,templeton2024scaling}. Building on this view, we hypothesize that effective audio LLM integration requires audio tokens to trigger the activation of sound related conceptual features inside the textual token embeddings. In other words, distinct auditory inputs should induce the corresponding linguistic concepts to become active in the text representation; for example, when the input contains a \emph{dog bark}, the features associated with the concept \emph{dog} should light up so the model can ground the auditory signal in language and answer queries such as \emph{Which animal sound is present?}. This hypothesis guides our architectural design: we seek the simplest pathway that reliably transmits audio cues into the text features that carry concepts.

A standard LLM layer consists of an attention submodule followed by a feed-forward network (FFN) submodule. Since attention mediates all inter-token interactions, it is the necessary pathway for audio to influence text, and we posit that it is also sufficient for text tokens to gather information from audio. Guided by this principle, we introduce \textbf{LAL} (Lightweight Audio LLM integration).

As in our baseline, a shared Q-Former produces a sequence of audio tokens and at each layer a small MLP projects these tokens into that layer's input space. Audio information is then injected into the attention block only through Keys and Values while Queries remain text only, so audio modulates the attention context of text tokens without passing through the feed-forward network.

Formally, let $H_l^t \in \mathbb{R}^{N_t \times d}$ denote the text hidden states at layer $l$ and $A \in \mathbb{R}^{N_a \times d_a}$ the Q-Former audio features. A per-layer projector $P_l:\mathbb{R}^{d_a}\!\to\!\mathbb{R}^{d}$ maps audio to the layer space,
\begin{equation}
\hat{A}_l = P_l(A) \in \mathbb{R}^{N_a \times d}
\end{equation}
and we concatenate text and audio along the token axis
\begin{equation}
S_l = \big[\, H_l^t \,;\, \hat{A}_l \,\big] \in \mathbb{R}^{(N_t+N_a)\times d}.
\label{eq_audio_insert}
\end{equation}
Queries are formed from \emph{text only} (see Figure~\ref{fig_pal_arch}(B)), while Keys and Values are computed from the concatenated sequence:
\begin{equation}
Q_l^t = H_l^t W_{Q,l},\qquad
K_l   = S_l   W_{K,l},\qquad
V_l   = S_l   W_{V,l}.
\label{eq_qkv_proj}
\end{equation}
The resulting LAL update for text tokens is
\begin{equation}
\tilde{H}_l^t = \mathrm{softmax}\!\left(\frac{Q_l^t K_l^\top}{\sqrt{d_k}}\right) V_l.
\label{eq_attn}
\end{equation}
after which $\tilde{H}_l^t$ proceeds through the FFN with the usual residual connections. In this way, audio cues shape the attention context seen by text tokens, aligning audio-evoked features with their linguistic counterparts and enabling effective cross–modal information transfer. Please note that in LAL we assign position IDs to both audio and text tokens to preserve token-order information, as detailed in Appendix~\ref{apx_lal_positionid} and Figure~\ref{fig_LAL_preserve}.

\textbf{Information Injection Dynamics. LAL is neither Lite PLITS nor Lite Flamingo.}
Beyond computational efficiency, LAL opens up a distinct information pathway. In PLITS, audio tokens are treated identically to text tokens: they are transformed layer by layer through causal self-attention and FFN non–linearities, causing their representations to drift from the original encoder output as they mix with the LLM's internal state. In contrast, LAL uses a dedicated MLP at each layer to project ``semantic-ready'' audio features directly into the appropriate abstraction for that layer. This preserves a direct link to the audio encoder's semantic output. For tasks that rely on explicit acoustic cues, such as sound event understanding (e.g., \emph{Which animal sound is heard?}), this projection-based injection can be more effective than the deeply transformed representations produced by PLITS.

When comparing to Flamingo, we note that although Flamingo also injects semantic level information without decoding audio tokens inside the LLM, the route by which this information influences text tokens is fundamentally different. In Flamingo, text tokens first attend to audio tokens in a dedicated cross attention module; the resulting signal is added to the text residual stream, and only then do the updated text states interact through the standard self attention layers of the LLM. In LAL, by contrast, audio representations are introduced directly into the same self attention operation as the text tokens, so text attends jointly to audio and text within a single attention computation. This produces a distinct information flow from Flamingo. We also note that LAL does not require the extra cross attention plus FFN adapter blocks used in Flamingo.

To summarize, LAL is similar to PLITS in that it performs in-context injection and allows text tokens to attend over both audio and text tokens, and it is similar to Flamingo in that it injects information that has not been fully decoded inside the LLM. However, it is architecturally distinct from both: LAL is neither a ``lite PLITS'' nor a ``lite Flamingo,'' but rather a new information pathway for integrating audio encoders with LLMs.

\textbf{LAL Integration with Frozen LLM FFN.} 
We also verify that LAL integration remains effective when the LLM's FFN blocks are frozen, 
with no significant loss in performance (refer to Appendix~\ref{apx_frozen_ffn}). 
This finding has important implications for reducing training cost, improving parameter efficiency, 
and preserving the pretrained knowledge of the LLM while enabling multimodal alignment. 
For clarity and consistency, however, our main experiments focus on the standard setting with trainable FFN blocks, 
and discussion of the frozen-FFN variant is limited to Appendix~\ref{apx_frozen_ffn}.

\textbf{Leveraging parametric versus contextual knowledge.} Here we posit how LAL \textit{efficiently} utilizes two types of knowledge inherent in pre-trained LLMs: (1) parametric knowledge, primarily embedded within the FFN layers as a result of extensive language pre-training, and (2) contextual knowledge, which is dynamically incorporated through attention mechanisms. We posit that audio as contextual information can effectively induce required concept activations in text token representations via attention-based modulation, without needing direct FFN processing of audio representations. Consequently, audio information indirectly accesses the LLM's parametric knowledge: the audio context "piggybacks" on text tokens, as attention mechanisms reconfigure these representations, which then engage relevant concept-related pathways during FFN processing. 

\subsubsection{Compute and memory efficiency.}
LAL is more compute- and memory-efficient than PLITS and Flamingo style integration, and the benefits become more pronounced with longer audio sequences. 
At a high level, the gains come from reducing the effective attention complexity and avoiding unnecessary routing of audio tokens through the feed-forward sublayers. Quantitative comparisons of memory usage and training throughput are reported in Figure \ref{fig_pal_comp}. 

In the following subsections, we present one-to-one comparisons between LAL and the PLITS baseline when applicable, and otherwise discuss properties specific to LAL.

\textbf{Attention Complexity:}

\textit{PLITS:} full causal attention over $N_a + N_t$ tokens with cost $\mathcal{O}\!\big((N_a{+}N_t)^2\big)$ 

\textit{LAL:} only text tokens issue queries; keys and values include audio and text, with cost $\mathcal{O}\!\big((N_a{+}N_t)N_t\big)$
  eliminating the $N_a^2$ term and all audio to audio interactions.

\textbf{Feedforward Routing:}

  \textit{PLITS:} audio tokens pass through attention and the feedforward sublayer in every block, increasing floating point operations and activation memory in proportion to $N_a$.
  
  \textit{LAL:} audio tokens do not enter the feedforward sublayer and only serve as keys and values for text queries, which reduces per layer floating point operations and activations stored for backpropagation.

\textbf{Scaling With Audio Length:}
Non text modalities in multimodal LLMs often yield far more tokens, and audio is no exception. As $N_a$ grows due to longer clips or denser tokenization, PLITS incurs a cost of $(N_a + N_t)^2$, so the $N_a^2$ term dominates. In contrast, LAL scales as $(N_a + N_t)N_t$, which is linear in $N_a$. Thus, the compute and memory gap widens with longer or more finely segmented audio. The feedforward savings in LAL also increase with $N_a$ as a larger share of tokens bypass the most expensive part of each block.

\textbf{Distinct from PEFT and LoRA:}
LAL is a core architectural modification, not a parameter-efficient fine-tuning (PEFT) method such as LoRA \citep{hu2022lora}. Techniques such as LoRA adjust how weights are adapted during training while keeping the forward compute pattern essentially the same at inference. In contrast, LAL changes how audio tokens participate in attention and feedforward routing, so its compute and memory savings apply at inference as well as during training.

\begin{table*}[t] 
\caption{Performance evaluation of the proposed efficient integration method \textbf{LAL} and SOTA integration \textbf{PLITS} across different base LLMs.
Evaluation follows the protocol of \citet{gong2024listen}. FI: Flamingo Integration, LFST: a Q-former-based approach that compresses features from SSLAM (fine-grained audio encoder) and CLAP (language-aligned audio encoder) into a single set of Q-former tokens (see Appendix~\ref{apx_lfst}), AC: AudioCaps, CL: Clotho, AS2M: AudioSet 2M.
\textsuperscript{\textdagger} indicates CIDEr and \textsuperscript{\textdaggerdbl} indicates SPICE.
Other metrics: accuracy (ESC-50, VocalSound), Mi-F1 (DCASE), and mAP (FSD, AudioSet). For evaluation methodology see Section~\ref{lal_eval_protocol} and for dataset details see Appendix~\ref{apx_evaluation}.}
\label{tab_lal_cls_cap} 

\centering 
\small 
\setlength{\tabcolsep}{2.0pt} 

\begin{tabular}{cccccccccccccc}
\hline
LLM & \multirow{2}{*}{PLITS} & \multirow{2}{*}{FI} &\multirow{2}{*}{LAL} & \multirow{2}{*}{LFST} & \multicolumn{5}{c}{Classification} & \multicolumn{4}{c}{Captioning} \\
\cline{6-14}
Backbone & & & & & ESC50 & DCASE & VS & FSD & AS2M & AC\textsuperscript{\textdagger} & CL\textsuperscript{\textdagger} & AC\textsuperscript{\textdaggerdbl} & CL\textsuperscript{\textdaggerdbl}\\
\hline
\multirow{4}{*}{Llama3.2-1B} & \ding{51} &\ding{55} & \ding{55} & \ding{55} & 64.45 & 37.69 & 51.57 & 25.23 & 9.08 & 0.59 & 0.34 & 16.30 & 10.96 \\
 & \ding{55} &\ding{55} & \ding{51} & \ding{55} & 76.70 & 40.97 & \textbf{60.87} & 31.44 & 11.83 & 0.66 & 0.38 & 16.97 & 11.87 \\
 & \ding{51} &\ding{55} & \ding{55} & \ding{51} & 84.10 & 45.28 & 57.59 & 42.49 & 14.74 & 0.70 & 0.39 & 17.90 & 11.82 \\
 & \ding{55} &\ding{51} & \ding{55} & \ding{51} & 84.95 & 43.95 & 55.44 & 41.27 & \textbf{15.0} & 0.69 & 0.39 & 17.09 & 11.91 \\
 & \ding{55} &\ding{55} & \ding{51} & \ding{51} & \textbf{87.40} & \textbf{46.23} & 56.03 & \textbf{43.91} & 14.74 & \textbf{0.72} & \textbf{0.42} & \textbf{18.08} & \textbf{12.58} \\
\hline
\multirow{4}{*}{Llama3.2-3B} & \ding{51} &\ding{55} & \ding{55} & \ding{55} & 70.40 & 40.62 & 61.40 & 28.88 & 10.84 & 0.63 & 0.35 & 16.81 & 11.35 \\
& \ding{55} &\ding{55} & \ding{51} & \ding{55} & 82.15 & 43.21 & \textbf{65.78} & 34.29 & 12.91 & 0.67 & 0.38 & 17.80 & 12.18 \\
 & \ding{51} &\ding{55} & \ding{55} & \ding{51} & 84.60 & 46.16 & 59.15 & 43.29 & 15.00 & 0.70 & 0.38 & 17.90 & 12.03 \\
 & \ding{55} &\ding{55} & \ding{51} & \ding{51} & \textbf{89.25} & \textbf{47.21} & 60.46 & \textbf{43.86} & \textbf{15.03} & \textbf{0.73} & \textbf{0.40} & \textbf{18.61} & \textbf{12.46} \\
\hline
\multirow{3}{*}{Qwen2.5-1.5B} & \ding{51} &\ding{55} & \ding{55} & \ding{55} & 68.00 & 37.57 & 56.45 & 27.87 & 9.56 & 0.63 & 0.38 & 16.63 & 11.74 \\
 & \ding{55} &\ding{55} & \ding{51} & \ding{55} & 70.85 & 38.79 & \textbf{59.20} & 28.53 & 10.28 & 0.63 & 0.38 & 16.65 & 11.44 \\
 & \ding{55} &\ding{55} & \ding{51} & \ding{51} & \textbf{87.80} & \textbf{45.52} & 56.73 & \textbf{43.26} & \textbf{13.92} & \textbf{0.73} & \textbf{0.41} & \textbf{18.45} & \textbf{12.20} \\
\hline
\end{tabular}
\end{table*}

\begin{table*}[h]
\centering
\caption{GPT-4 evaluation of \textsc{LAL} and \textsc{PLITS} on the CompA-R benchmark~\citep{ghosh2024gama}. LFST: a Q-former-based approach that compresses features from SSLAM (fine-grained audio encoder) and CLAP (language-aligned audio encoder) into a single set of Q-former tokens (see Appendix~\ref{apx_lfst}). A text only GPT-4 judge scores the model outputs; see~\citet{ghosh2024gama} for the detailed prompt.}
\small

\setlength{\tabcolsep}{2.0pt}
 \begin{tabular}{cccccccc}
 \hline
{PLITS} & {LAL} & {LFST}  &  Helpfulness & Clarity & Correctness & Depth & Engagement \\
\hline
\ding{51} & \ding{55} & \ding{51} & \textbf{3.86} & \textbf{4.74} & \textbf{3.84} & 2.86 & 2.99 \\
\ding{55} & \ding{51} & \ding{51}  &3.85 & 4.70 & 3.82 & \textbf{2.88} & \textbf{3.01} \\ 
\hline
\end{tabular}
\label{tab_lal_reasoning}
\label{compr}
\end{table*}


\subsection{Extending to Speech and Building Unified General Audio-Music-Speech Understanding LLMs}
\label{sec_pal}

In this section, we expand our framework to encompass a unified model capable of speech, sound, and music understanding.

\begin{figure}[bt] 
\centering
  \includegraphics[width=0.60\linewidth]{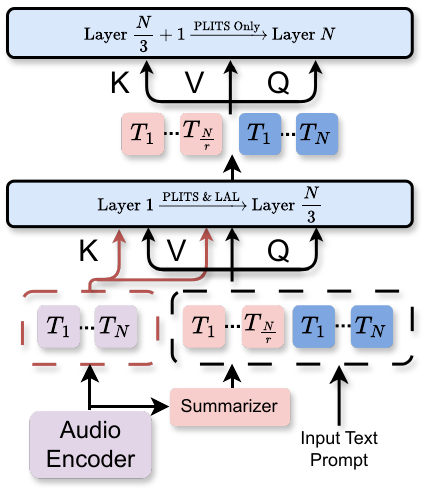} 
\caption{Overview of \textbf{PAL}. A unified audio encoder produces full-resolution audio tokens, which are (i) summarized into a compact set of PLITS tokens that are prepended and processed through all LLM layers, and (ii) injected via LAL into the \emph{initial one-third} of LL layers. Beyond the first third of layers, PAL relies only on the PLITS summary tokens.}
\label{fig_pal_arch_final}
\end{figure}

Although PLITS and LAL are each reasonably effective on their own, we investigate whether combining them yields additional benefits for this broader unified setting. Concretely, we use LAL to inject full-resolution contextual information and PLITS to provide compact summary representations, improving efficiency. We hypothesize that, in the \textbf{PAL} hybrid setup, where \textbf{PLITS} and \textbf{LAL} provide \emph{competing pathways} for audio information flow, \textbf{LAL} is most valuable in the early transformer layers, where preserving fine-grained acoustic detail is critical. In contrast, \textbf{PLITS} becomes more effective in later layers as intermediate representations become increasingly aligned with text tokens. In \textbf{standalone} PLITS-only or LAL-only integration, this pathway competition does not arise: each method serves as the model's only pathway for transferring all necessary audio information into the LLM. Accordingly, we restrict LAL integration to the first third of the model's layers. We call this hybrid architecture \textbf{PAL} (\textbf{P}robing \textbf{A}udio encoders via \textbf{L}LMs), which enables the language model to efficiently probe audio features. The evolution of PAL from PLITS and LAL, along with a corresponding analysis, is presented in Appendix~\ref{apx_pal_evolution}.

Architecturally, we use a unified audio encoder capable of understanding speech, music, and sound (e.g., AFWhisper~\cite{goel2025audio}). To balance efficiency with the benefits of PLITS, we construct two parallel views of the resulting token sequence:

\textbf{1. Compact Summary Tokens (PLITS Pathway):} We derive a compact set of summary tokens by applying a one-dimensional convolution with stride $r$ along the time axis, reducing the token count by a factor of $r$ (we use $r=3$ throughout this work). These summary tokens are treated as PLITS tokens: they are mapped into the LLM input space using an MLP, prepended to the text tokens, and processed as full tokens through all LLM layers.

\textbf{2. Full-Resolution Tokens (LAL Pathway):} We retain the complete token sequence produced by the audio encoder (AFWhisper) for LAL integration. At each LLM layer, we project these tokens into the LLM embedding space using an MLP. Audio information is then injected into the attention block by supplying \emph{both} the full-resolution tokens and the PLITS-integrated summary tokens as Keys and Values, while Queries are issued by the text tokens and the summary tokens.

To preserve temporal ordering, we interleave the tokens in the attention mechanism (specifically in the Keys and Values) such that, for every span of $r$ LAL tokens, the corresponding summary token follows its source tokens (see Figure~\ref{fig_audio_uni_encoder}). Concretely, the ordering of Keys and Values in the attention module follows:
\begin{equation}
\mathbf{z}
= (\ell_1,\; \ell_2,\; \ldots,\; \ell_r,\; p_1,\; \ell_{r+1},\; \ldots,\; \ell_{2r},\; p_2,\; \ldots),
\label{eq:audio_interleave}
\end{equation}
where $\ell_i$ denotes an LAL token and $p_j$ denotes a summary token integrated via PLITS. Please refer to Appendix~\ref{apx_pal_token_ordering} for more details of PAL token ordering. This maintains the alignment between summary tokens and their underlying fine-grained audio context. To maximize efficiency without sacrificing performance, LAL integration is applied primarily in the initial one-third of the network's layers.  An overview of the proposed PAL audio-LLM is shown in Figure~\ref{fig_pal_arch_final}, and a detailed ablation study motivating these design choices is provided in Appendix~\ref{apx_pal_evolution}.

\section{Experiments and Results}
Our experiments follow the methodology progression. We first study \textbf{LAL} as an efficient alternative to the standard \textbf{PLITS} integration on \emph{general audio understanding}, establishing that attention-only injection can transfer audio-event information effectively while reducing compute and memory. We then move to a \emph{unified} setting that requires strong performance on \textbf{speech, sound, and music} simultaneously, and evaluate {LAL}, {PLITS}, and the proposed hybrid {PAL}.

\subsection{General Audio Understanding: LAL vs.\ PLITS}
\label{lal_eval_protocol}

\paragraph{Training pipeline.}
This experimental setup consists following stages:
\textbf{connector pretraining,} where only the audio-LLM connector is trained while the LLM and audio encoder are kept frozen; and
\textbf{joint training}, where the connector and the LLM are trained together.
The audio encoders remain frozen throughout.
We use two large-scale instruction-tuning corpora: OpenAQA \citep{gong2024listen} for these 2 stages. 

Additionally, to support open-ended question answering and reasoning, we extend training with additional open-ended samples from OpenAQA (\textit{Stage~3}) and the dedicated reasoning set CompA-R (\textit{Stage~4}). Full details on hyper-parameters are provided in Appendix~\ref{apx_general_audio_hyperparam}.

As the audio encoder, we use LFST, a Q-former-based compression method that combines features from SSLAM~\cite{alex2025sslam} (a fine-grained audio encoder) and CLAP~\cite{elizalde2023clap} (a language-aligned audio encoder) into a single set of Q-former tokens (see Appendix~\ref{apx_lfst}). As the base LLM, we use Llama-3.2-1B Instruct \citep{grattafiori2024llama}. To study scaling and transfer across model families, we additionally report results with a larger Llama-3.2-3B \citep{grattafiori2024llama} backbone and with Qwen2.5~1.5B Instruct \citep{qwen2.5}.

\paragraph{Evaluation tasks.}
To quantify how effectively each integration method transfers audio-event information from the frozen encoder into the LLM’s latent space, we evaluate on three complementary settings: classification, captioning, and reasoning. For \textbf{classification}, we follow the LTU protocol \citep{gong2024listen} and measure semantic similarity between the model’s textual prediction and the ground-truth label by embedding both using \texttt{gpt-text-embedding-ada}, reporting similarity-based accuracy. For \textbf{captioning}, we evaluate on standard audio captioning benchmarks and report CIDEr and SPICE. For \textbf{reasoning}, we use \texttt{CompA-R-test} under the evaluation procedure of \citet{ghosh2024gama}, where a text-only GPT-4 judge rates generated answers along five dimensions: \emph{Helpfulness}, \emph{Clarity}, \emph{Correctness}, \emph{Depth}, and \emph{Engagement}.

\paragraph{Results.}
We present results in two tiers to clearly separate architectural contribution from comparisons to prior systems.
First, we provide a controlled comparison between \textbf{LAL}, and state-of-the art \textbf{PLITS}, and also Flamingo-style: classification and captioning in Table~\ref{tab_lal_cls_cap}, and reasoning in Table~\ref{tab_lal_reasoning}. These results show that LAL matches or improves performance while offering substantial efficiency benefits (throughput and memory), consistent with the analysis in Figure~\ref{fig_pal_comp}.

Second, we compare against prior published methods for both (i) classification/captioning and (ii) reasoning; due to space constraints these expanded comparisons are deferred to Appendix~\ref{apx_cap_cls_detailed} (Tables~\ref{tab_lal_prior} and~\ref{tab_lal_prior_compar}).

\subsection{Unified Speech, Music, and Sound Understanding: LAL vs.\ PLITS vs.\ PAL}
\label{pal_eval_protocol}

\paragraph{Training protocol.}
In the unified setting, we train the audio-LLM to handle speech, general audio, and music. We follow a \textbf{two-stage} training pipeline:
(i) \textbf{connector pretraining}, where only the audio-LLM connector is trained while the LLM and audio encoder are kept frozen; and
(ii) \textbf{joint training}, where the connector and the LLM are trained together.
The audio encoder remains frozen throughout.

For Stage~1, we construct a mixture from the OpenAQA Stage~1 general-audio set and augment it with the OpenASQA\citep{gong2023joint} Stage~1 split for speech understanding. For Stage~2, we use a curated multi-domain reasoning instruction-tuning corpus, namely a 6M subset of AudioSkills \citep{goel2025audio} due to the unavailability of original audio files for some source datasets.

As the audio encoder, we use AFWhisper, a unified encoder for speech, music, and sound, followed by an MLP that projects its representations into the LLM embedding space for both PLITS and LAL. For this setting, we use Llama-3.2-3B \citep{grattafiori2024llama} as the LLM backbone for all experiments.

\paragraph{Evaluation protocol.}
We evaluate unified audio-speech-music understanding on three benchmarks. \textbf{MMAU} \citep{sakshi2024mmau} is a multiple-choice audio QA benchmark spanning \emph{speech, sound, and music} and covering a broad set of skills; we report accuracy overall and by domain on the official \textbf{test-mini} and \textbf{test} splits. \textbf{MMAR} \citep{ma2025mmar} targets \emph{deep audio reasoning} with expert-authored multiple-choice questions sourced from real-world videos and organized into four reasoning layers (\emph{Signal, Perception, Semantic, Cultural}); we report accuracy following the benchmark protocol. \textbf{MMSU} \citep{wang2025mmsu} focuses on spoken-language understanding and reasoning across diverse speech phenomena (e.g., phonetics, prosody, semantics, and paralinguistics); we report overall accuracy along with perception- versus reasoning-oriented breakdowns.

\paragraph{Results.}
Using \textbf{PLITS} as the baseline, we find that our proposed \textbf{LAL} integration is \emph{competitive} despite being substantially more efficient.
On \textbf{MMAU}, LAL-3B slightly improves overall \textbf{test} accuracy over PLITS-3B (63.24 vs.\ 62.91), with clear gains on the \textbf{speech} subset (50.07 vs.\ 46.48), while remaining broadly comparable across domains.
On \textbf{MMSU}, LAL-3B also matches or exceeds the baseline overall (41.22 vs.\ 40.12), indicating that injecting audio via attention alone can preserve strong unified speech reasoning and perception.
Notably, the longer-trained \textbf{LAL-3B (LT)} further closes the gap and strengthens performance, reaching 42.76 (Avg-All) on MMSU and improving over PLITS on \textbf{MMAR} (43.50 vs.\ 41.10).
This trend is consistent with LAL’s lower training and inference overhead: with comparable compute budgets, LAL can be trained longer (or applied to longer audio at inference), translating efficiency into accuracy gains.

Finally, the proposed hybrid \textbf{PAL} achieves the best performance across all benchmarks, delivering large improvements on \textbf{MMAU} (70.20 / 68.20 on test-mini / test) and the highest accuracy on \textbf{MMAR} (45.40) and \textbf{MMSU} (46.18).
These results validate that combining full-resolution LAL context for early layers with compact PLITS summary tokens yields the strongest unified speech-music-sound understanding, with especially strong gains on speech-intensive evaluation and better efficiency compared to PLITS.

\begin{table*}[t]
\centering
\caption{Performance summary on MMAU-v05.15.25 and MMAR benchmarks across different integration architectures. Note that values are missing (-) for LAL+ 3B as the MMAU-test online challenge was closed at the time of evaluation. A detailed comparative analysis against existing state-of-the-art audio-LLMs is available in Appendix~\ref{apx_mmau_mmar_mmsu}. $^{\dagger}$ \textit{LT indicates the same model trained for more steps (no architectural changes); excluded from bolding to ensure a fair comparison.}}
\label{tab_mmau_mmar}
\small
\setlength{\tabcolsep}{8pt}
\begin{tabular}{llcccccccc}
\hline
\multirow{2}{*}{\textbf{Benchmark}} & \multirow{2}{*}{\textbf{Model}} & \multicolumn{2}{c}{\textbf{Sound}} & \multicolumn{2}{c}{\textbf{Music}} & \multicolumn{2}{c}{\textbf{Speech}} & \multicolumn{2}{c}{\textbf{Total Acc.}} \\
 &  & mini & test & mini & test & mini & test & mini & test \\ \hline
\multirow{4}{*}{\textbf{MMAU}} 
 &  PLITS-3B & 75.68 & 72.03 & 70.96 & 69.63 & 46.25 & 46.48 & 64.30 & 62.91 \\
 & LAL-3B & 73.57 & 69.53 & 67.66 & 69.67 & 47.15 & 50.07 & 62.80 & 63.24 \\
 & \cellcolor{gray!15}LAL-3B (LT)$^{\dagger}$ & \cellcolor{gray!15}75.08 & \cellcolor{gray!15} -  & \cellcolor{gray!15}70.66 & \cellcolor{gray!15} - & \cellcolor{gray!15}48.05 & \cellcolor{gray!15} - & \cellcolor{gray!15}64.60 & \cellcolor{gray!15} - \\
 & PAL-3B & \textbf{78.08} & \textbf{74.60} & \textbf{73.05} & \textbf{71.30} & \textbf{59.46} & \textbf{58.36} & \textbf{70.20} & \textbf{68.20} \\ \hline
\multirow{4}{*}{\textbf{MMAR}} 
& PLITS-3B & \multicolumn{2}{c}{38.79} & \multicolumn{2}{c}{40.29} & \multicolumn{2}{c}{37.41} & \multicolumn{2}{c}{41.10} \\
 & LAL-3B & \multicolumn{2}{c}{38.18} & \multicolumn{2}{c}{\textbf{42.72}} & \multicolumn{2}{c}{35.37} & \multicolumn{2}{c}{38.90} \\
 & \cellcolor{gray!15}LAL-3B (LT)$^{\dagger}$ & \multicolumn{2}{c}{\cellcolor{gray!15}47.88} & \multicolumn{2}{c}{\cellcolor{gray!15}43.69} & \multicolumn{2}{c}{\cellcolor{gray!15}38.10} & \multicolumn{2}{c}{\cellcolor{gray!15}43.50} \\
 & PAL-3B & \multicolumn{2}{c}{\textbf{44.85}} & \multicolumn{2}{c}{41.75} & \multicolumn{2}{c}{\textbf{43.54}} & \multicolumn{2}{c}{\textbf{45.40}} \\ \hline
\end{tabular}
\end{table*}
\begin{table*}[t]
\centering
\caption{Performance Comparison on the MMSU Benchmark across perception and reasoning
dimensions in Semantics (Seman.), Phonology (Phono.), and Paralinguistics (Para.) domains. A more detailed comparative analysis against existing state-of-the-art audio-LLMs is available in Appendix~\ref{apx_mmau_mmar_mmsu}.}
\label{tab_mmsu}
\small
\setlength{\tabcolsep}{6pt}
\begin{tabular}{llccccccccc}
\hline
\multirow{2}{*}{\textbf{Benchmark}} & \multicolumn{1}{c}{} & \multicolumn{4}{c}{\textbf{Perception}} & \multicolumn{4}{c}{\textbf{Reasoning}} & \textbf{Avg} \\

& \multicolumn{1}{c}{\multirow{-2}{*}{\textbf{Model}}} & Seman. & Phono.  &Para.& Avg &  Seman. & Phono.  &Para.& Avg  & \textbf{All} \\
\hline
\multirow{4}{*}{\textbf{MMSU}}  &PLITS-3B &24.09 &	31.98 &	34.75 &	31.12 &	49.46	 &52.3	 &42.99	 &49.71 &	40.12 \\
 &LAL-3B  & 26.61	&33.8	&31.19	&31.01&	50.00	&54.35	&\textbf{52.54}	&52.11&	41.22  \\
 &\cellcolor{gray!15}LAL-3B (LT)$^{\dagger}$ & \cellcolor{gray!15}31.34 &\cellcolor{gray!15}34.33 &\cellcolor{gray!15} 27.13 &\cellcolor{gray!15} 30.78 &\cellcolor{gray!15} 53.97 &\cellcolor{gray!15} 58.75 & \cellcolor{gray!15}51.34 &\cellcolor{gray!15} 55.54 &\cellcolor{gray!15} 42.76 \\ 
& PAL-3B & \textbf{29.76}&	\textbf{37.65}	&\textbf{37.52}	&\textbf{35.66}&	\textbf{58.12}&	\textbf{59.16}&	49.85&	\textbf{57.4}	& \textbf{46.18} \\ \hline

\end{tabular}
\end{table*}


\section{Conclusion}
We introduce LAL, which injects audio only through attention keys and values and skips feedforward processing for audio tokens. 
This reduces attention interactions and activations, yielding up to about $60\%$ lower memory usage and up to about $190\%$ higher training throughput, with performance comparable to PLITS, the state of the art baseline integration for classification, captioning, and reasoning tasks. We also propose PAL, an hybrid integration that uses LAL both PLITS for efficient audio-LLM that understand general audio and speech. LAL is a core architectural change rather than a parameter efficient fine tuning method, so the efficiency gains hold at inference and during training. For future work, we plan to scale to larger LLMs, use higher quality instruction data to improve reasoning, and explore streaming and long context audio.

\section*{Impact Statement}
This work introduces LAL and PAL, efficient audio-LLM integration methods that improve scalability while matching or exceeding prior approaches (e.g., PLITS/Flamingo) on broad audio understanding tasks. Audio-LLMs may enable privacy-invasive monitoring (e.g., surveillance of speech or ambient audio without consent), and can reflect dataset biases across languages, accents, demographics, or musical genres. They may also produce confident but incorrect interpretations of ambiguous audio, which is risky in high-stakes settings. We encourage responsible deployment: obtain consent and protect privacy for real-world audio, avoid high-stakes use without human oversight, evaluate across diverse conditions and populations, and follow appropriate data and licensing practices. Overall, LAL/PAL’s efficiency gains can broaden access to audio understanding, but should be paired with transparency and safeguards to mitigate misuse and unintended harms.

\section*{Acknowledgments}

This research was supported by the EPSRC and BBC Prosperity Partnership ``AI4ME: Future Personalized Object Based Media Experiences Delivered at Scale Anywhere'' (EP/V038087/1). For part of the experiments, we used resources provided by the EuroHPC Joint Undertaking, which granted this project access through a EuroHPC Development Access call to the LEONARDO EuroHPC supercomputer hosted by CINECA (Italy) and to the resources of the LEONARDO consortium. The authors also acknowledge the use of resources provided by the Isambard-AI National AI Research Resource (AIRR)~\cite{mcintoshsmith2024isambardaileadershipclasssupercomputer}. Isambard-AI is operated by the University of Bristol and is funded by the UK Government's Department for Science, Innovation and Technology (DSIT) via UK Research and Innovation; and the Science and Technology Facilities Council [ST/AIRR/I-A-I/1023].




\nocite{langley00}

\bibliography{example_paper}
\bibliographystyle{icml2026}

\newpage
\appendix
\onecolumn


\section{Evolution of PAL from LAL and PLITS}
\label{apx_pal_evolution}

In this section, we motivate the architectural choices behind \textbf{PAL}, our hybrid integration that combines \textbf{PLITS} and \textbf{LAL}.
Across Tables~\ref{tab_mmau_mmar}, \ref{tab_mmsu}, \ref{tab_lal_cls_cap}, and \ref{tab_lal_reasoning}, \textbf{LAL} emerges as an efficient alternative to PLITS: it maintains comparable performance under the same training setup, and with longer training it matches or surpasses the PLITS baseline.
At the same time, PLITS and LAL expose the LLM to audio information through fundamentally different pathways, PLITS inserts projected audio tokens into the LLM input stream, whereas LAL injects audio only through attention at selected layers.
Motivated by this complementarity, we investigate whether a \emph{hybrid} design can combine the strengths of both approaches while being efficient.

As described in Section~\ref{sec_pal}, PLITS tokens are propagated through \emph{all} LLM layers via the standard token sequence, while LAL provides the flexibility to introduce full-resolution audio context at arbitrary layers of the LLM.
We therefore study how the choice of LAL insertion depth affects performance, and to what extent each integration pathway contributes to the final predictions.

~\paragraph{Masking PLITS and LAL pathways.} To this end, we first construct a PAL variant that applies LAL to \emph{all} LLM layers (denoted \textbf{PAL-All}). We then perform inference-time ablations by masking either the LAL pathway or the PLITS summary tokens, isolating their individual contributions.
The resulting performance trends are reported in Table~\ref{tab_mmau_mmar_pal_evolution} and Table~\ref{tab_mmsu_pal_evolution}.

Tables~\ref{tab_mmau_mmar_pal_evolution} and~\ref{tab_mmsu_pal_evolution} demonstrate that \textbf{PAL-All} benefits from \emph{both} integration pathways, and that each pathway contributes complementary information.
On \textbf{MMAU}, masking either LAL or PLITS leads to a clear drop in overall accuracy (PAL-All-3B: 66.26 test vs.\ 61.52 with LAL masked and 61.36 with PLITS masked), indicating that neither pathway alone recovers the full PAL-All performance.
The domain breakdown further highlights complementarity: removing \textbf{LAL} hurts \emph{music} more (70.03$\rightarrow$66.17 on MMAU-test), whereas removing \textbf{PLITS} hurts \emph{speech} more (54.46$\rightarrow$49.90 on MMAU-test), consistent with LAL providing richer fine-grained acoustic context and PLITS summary tokens providing a strong high-level signal aligned with language modeling in the hybrid PAL setting.

A similar pattern appears on \textbf{MMAR}, where disabling either pathway degrades performance, but masking \textbf{PLITS} is more damaging overall (44.4$\rightarrow$39.2) than masking \textbf{LAL} (44.4$\rightarrow$41.6), with the largest drops on \emph{sound} and \emph{speech}.
On \textbf{MMSU}, both ablations again reduce accuracy (43.34$\rightarrow$40.34 with LAL masked; 43.34$\rightarrow$39.88 with PLITS masked), and the splits suggest a mild specialization: LAL masking impacts \emph{perception} more (33.53$\rightarrow$29.81), while PLITS masking impacts \emph{reasoning} slightly more (53.8$\rightarrow$50.87).
Overall, these ablations support that LAL and PLITS provide complementary cues inside PAL-All, and that combining them yields a stronger unified model than relying on either pathway alone.

\begin{figure*}[t]
\centering

\newcommand{\panelH}{0.24\textheight}

\begin{minipage}[t]{0.49\textwidth}
  \centering
  \includegraphics[height=\panelH,width=\linewidth,keepaspectratio]{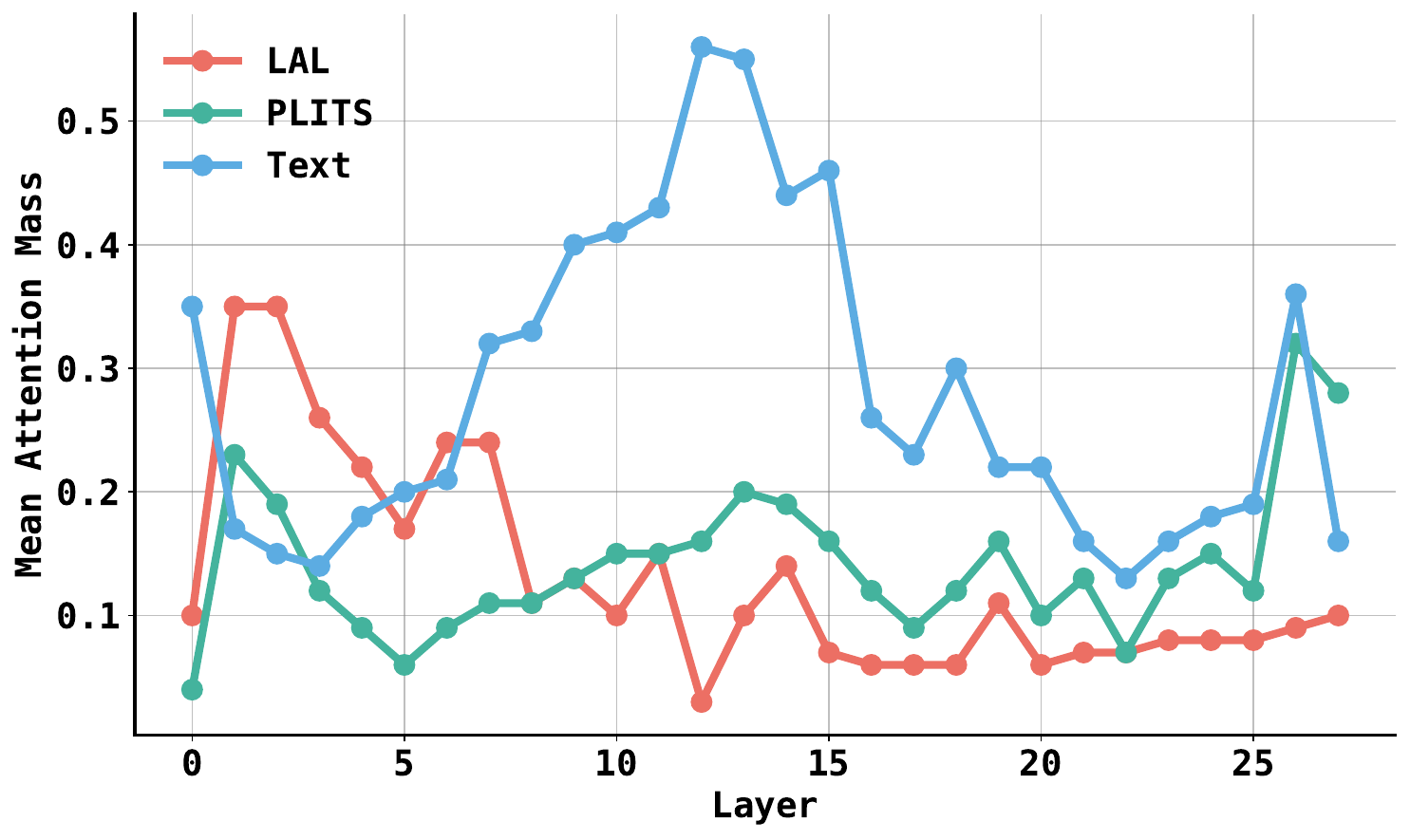}\\[-0.3em]
  {\small (a) Sound}
\end{minipage}\hfill
\begin{minipage}[t]{0.49\textwidth}
  \centering
  \includegraphics[height=\panelH,width=\linewidth,keepaspectratio]{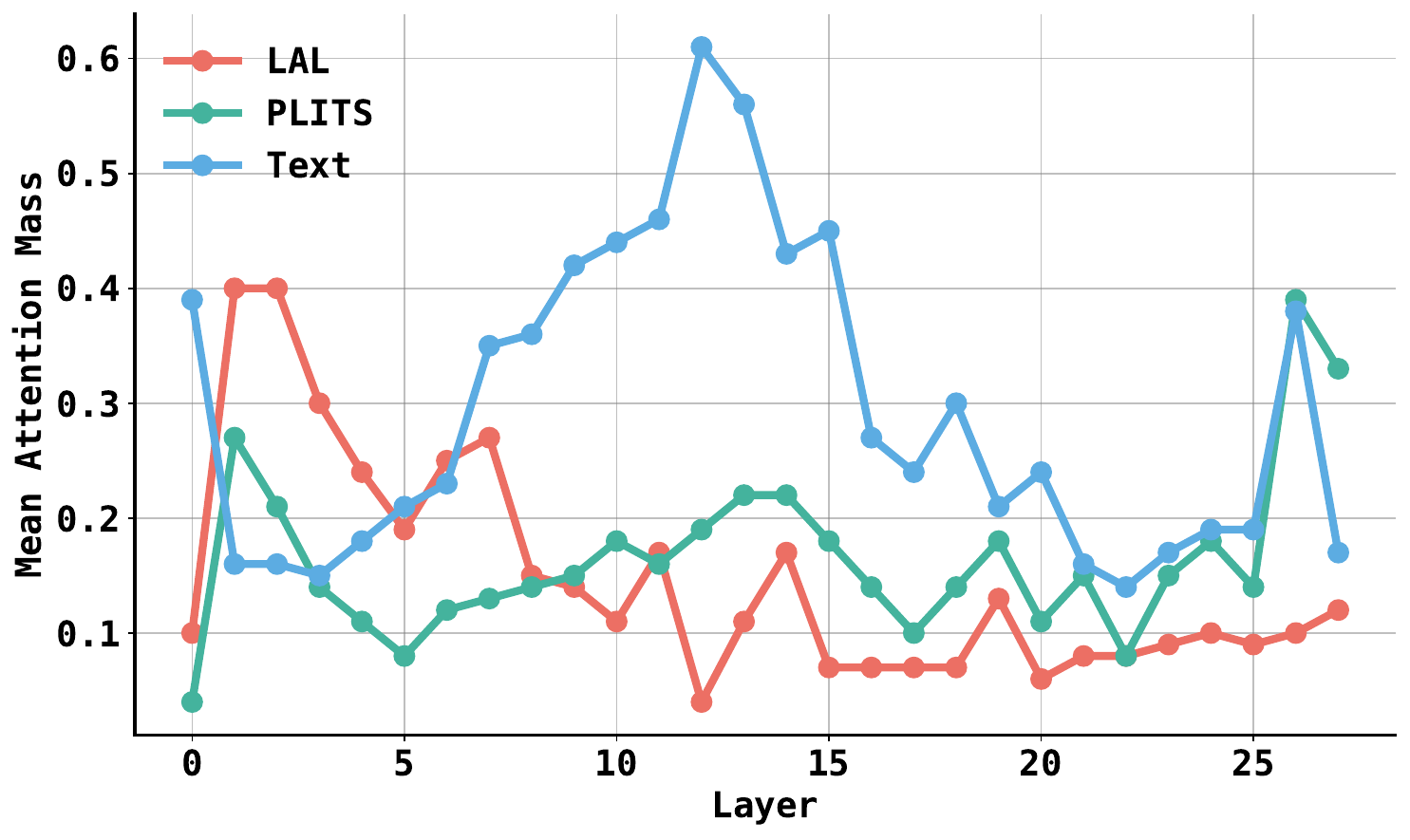}\\[-0.3em]
  {\small (b) Music}
\end{minipage}

\vspace{0.6em}

\begin{minipage}[t]{0.49\textwidth}
  \centering
  \includegraphics[height=\panelH,width=\linewidth,keepaspectratio]{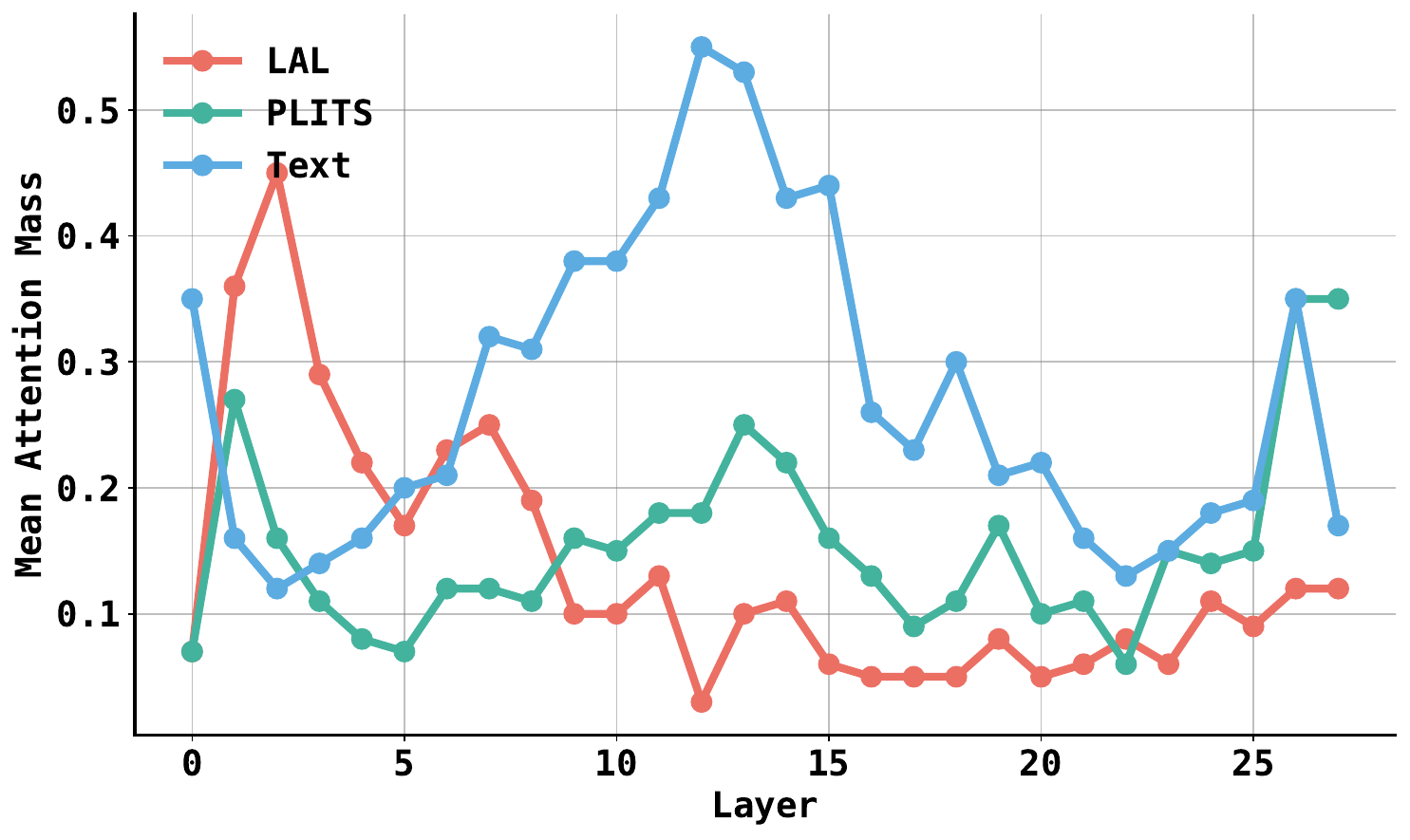}\\[-0.3em]
  {\small (c) Speech}
\end{minipage}

\vspace{-0.4em}
\caption{Mean attention mass per LLM layer on MMAR, decomposed by token group (LAL, PLITS, text), for (a) sound, (b) music, and (c) speech.}

\label{fig_attn_contribution}
\end{figure*}

\paragraph{Attention contribution analysis.}
To better understand how the model uses each pathway across depth, we compute the layer-wise mean attention mass assigned to token groups (LAL tokens, PLITS summary tokens, and text/system tokens) in \textbf{PAL-All-3B} and visualize the results on the sound/music/speech subsets of MMAR in Figure~\ref{fig_attn_contribution}.
Across all three domains, attention to \textbf{LAL} is most prominent in the early layers and decreases in deeper blocks, while attention to \textbf{PLITS} remains relatively lower early on and becomes more competitive toward later layers.
This supports our design hypothesis that full-resolution LAL features are most beneficial for early acoustic processing, whereas PLITS summaries better align with later, more text-oriented reasoning.
Guided by this observation, the final \textbf{PAL-3B} restricts LAL injection to the \emph{initial one-third} of layers, which yields consistent gains over PAL-All on MMAU/MMAR/MMSU (Tables~\ref{tab_mmau_mmar_pal_evolution} and~\ref{tab_mmsu_pal_evolution}).

\section{PAL: Token ordering inside Attention of PAL}
\label{apx_pal_token_ordering}
\begin{figure}[h]
\begin{center}
\includegraphics[width=0.7\linewidth]{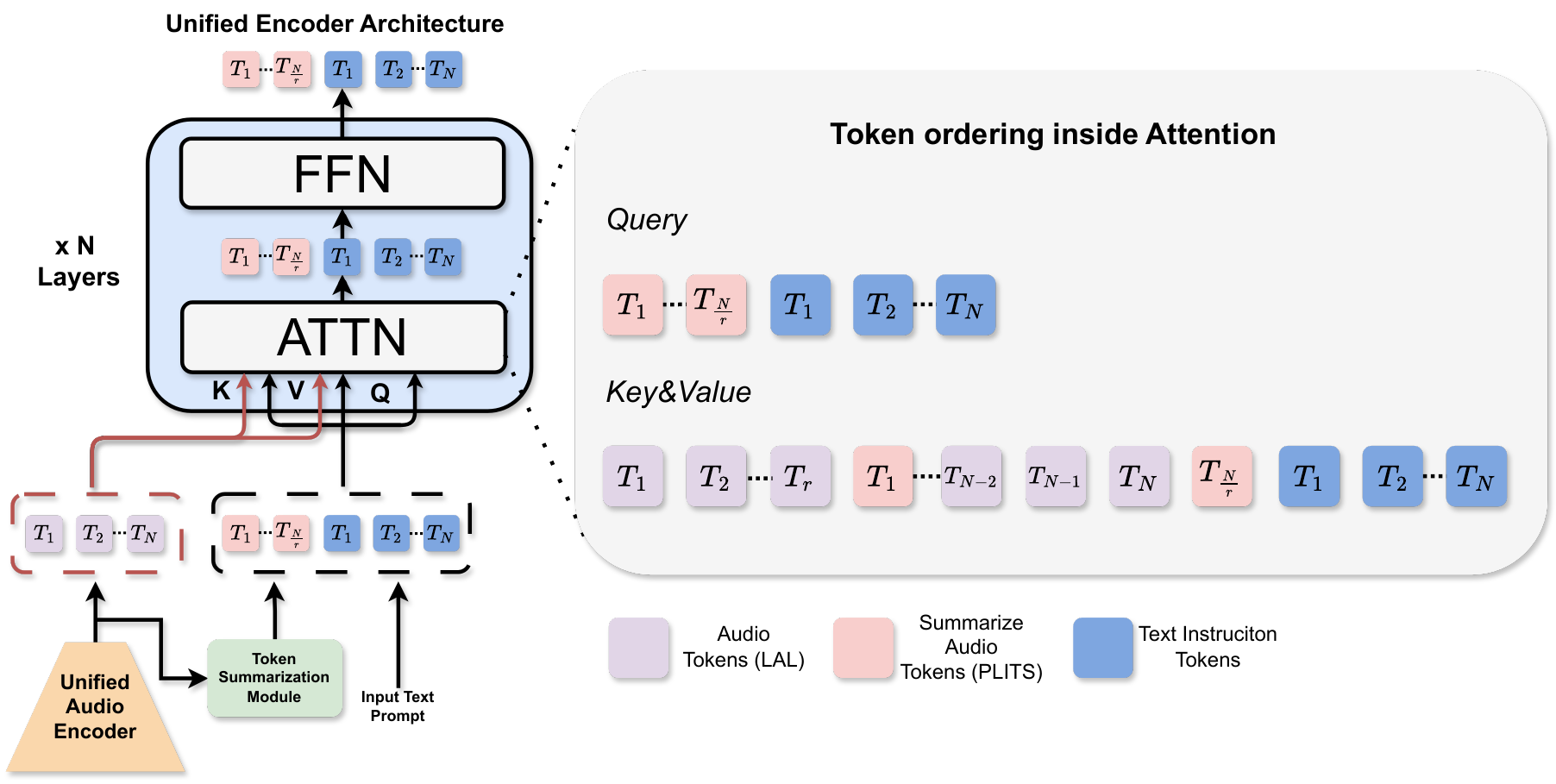}
\end{center}
\caption{Overview of PAL integration with unified audio encoder. The unified audio encoder processes input audio and produces audio tokens that are split into two paths. Tokens following LAL integration are $[T_1^{LAL}, \ldots, T_N^{LAL}]$, while token summarization reduces them by factor $r$ for PLITS integration producing $[T_1^{PLITS}, \ldots, T_{N/r}^{PLITS}]$. Purple tokens represent LAL audio tokens, red tokens represent summarized PLITS audio tokens, and blue tokens represent text tokens $[T_1^{text}, \ldots, T_N^{text}]$. In the attention mechanism, the query is ordered as $[T^{PLITS}, T^{text}]$ from PLITS integration, while the key and value tensors are ordered as the interleaving of audio tokens $T^{LAL}$ with their corresponding summarized tokens $T^{PLITS}$, then text tokens $T^{text}$(Section ~\ref{sec_pal}).}
\label{fig_audio_uni_encoder}
\end{figure}

In PAL, the Query contains audio PLITS tokens and text tokens, while the Key and Value contain the full sequence of text tokens, audio PLITS tokens, and injected audio LAL tokens. The tokens are arranged in an interleaved sequence where each summarized PLITS token is positioned immediately after its corresponding injected LAL tokens (non-summarized audio), followed by the LAL tokens of the next audio segment (as shown in Figure \ref{fig_audio_uni_encoder}). This interleaving enables the summarized PLITS tokens to attend to the injected LAL tokens from their own segment through the causal attention mechanism, while preventing attention to LAL tokens from subsequent segments.

\begin{table*}[t]
\centering
\caption{Ablation on MMAU-v05.15.25 and MMAR to quantify the contributions of the two PAL pathways.
\textbf{PAL-All-3B} applies LAL injection at \emph{all} LLM layers together with PLITS summary tokens.
\textbf{PAL-All-3B-LAL Masked} disables the LAL pathway at inference (PLITS-only within the PAL-All architecture),
while \textbf{PAL-All-3B-PLITS Masked} removes the PLITS summary tokens (LAL-only).
\textbf{PAL-3B} is the final PAL design, where LAL is applied only to the \emph{initial one-third} of LLM layers (Section~\ref{sec_pal}).}

\label{tab_mmau_mmar_pal_evolution}
\small
\setlength{\tabcolsep}{8pt}
\begin{tabular}{llcccccccc}
\hline
\multirow{2}{*}{\textbf{Benchmark}} & \multirow{2}{*}{\textbf{Model}} & \multicolumn{2}{c}{\textbf{Sound}} & \multicolumn{2}{c}{\textbf{Music}} & \multicolumn{2}{c}{\textbf{Speech}} & \multicolumn{2}{c}{\textbf{Total Acc.}} \\
 &  & mini & test & mini & test & mini & test & mini & test \\ \hline
\multirow{4}{*}{\textbf{MMAU}} 
 &  PAL-All-3B &76.28 &  73.87 &  69.76 & 70.03 & 49.25 & 54.46 & 65.10 & 66.26 \\
 & PAL-All-3B-LAL Masked & 69.67 & 67.17 & 67.96 & 66.17 & 49.55 & 50.86 & 62.4 & 61.52 \\
 & PAL-All-3B-PLITS Masked & 69.07 & 66.60 & 66.17 & 67.17 & 45.65 &49.90 & 60.3 & 61.36 \\ \cline{2-10}
& PAL-3B & \textbf{78.08} & \textbf{74.60} & \textbf{73.05} & \textbf{71.30} & \textbf{59.46} & \textbf{58.36} & \textbf{70.20} & \textbf{68.20} \\ \hline
\multirow{4}{*}{\textbf{MMAR}} 
& PAL-All-3B & \multicolumn{2}{c}{\textbf{46.67}} & \multicolumn{2}{c}{44.17} & \multicolumn{2}{c}{40.82} & \multicolumn{2}{c}{44.4} \\
 & PAL-All-3B-LAL Masked & \multicolumn{2}{c}{42.42} & \multicolumn{2}{c}{\textbf{41.75}} & \multicolumn{2}{c}{40.48} & \multicolumn{2}{c}{41.6} \\
 & PAL-All-3B-PLITS Masked & \multicolumn{2}{c}{40.00} & \multicolumn{2}{c}{39.32} & \multicolumn{2}{c}{36.73} & \multicolumn{2}{c}{39.2} \\ \cline{2-10}
  & PAL-3B & \multicolumn{2}{c}{44.85} & \multicolumn{2}{c}{\textbf{41.75}} & \multicolumn{2}{c}{\textbf{43.54}} & \multicolumn{2}{c}{\textbf{45.40}} \\ \hline
\end{tabular} 
\end{table*}
\begin{table*}[t]
\centering
\caption{Ablation on MMSU across perception and reasoning
dimensions in Semantics (Seman.), Phonology (Phono.), and Paralinguistics (Para.) domains to isolate the contributions of PLITS summary tokens and LAL injection in PAL.
\textbf{PAL-All-3B} uses both pathways with LAL connected to \emph{all} LLM layers.
\textbf{PAL-All-3B-LAL Masked} disables LAL at inference (retaining only PLITS summary tokens),
and \textbf{PAL-All-3B-PLITS Masked} removes PLITS summary tokens (retaining only LAL).
\textbf{PAL-3B} denotes the final PAL configuration, where LAL is applied only to the \emph{initial one-third} of LLM layers (Section~\ref{sec_pal}).}

\label{tab_mmsu_pal_evolution}
\small
\setlength{\tabcolsep}{6pt}
\begin{tabular}{llccccccccc}
\hline
\multirow{2}{*}{\textbf{Benchmark}} & \multicolumn{1}{c}{} & \multicolumn{4}{c}{\textbf{Perception}} & \multicolumn{4}{c}{\textbf{Reasoning}} & \textbf{Avg} \\

& \multicolumn{1}{c}{\multirow{-2}{*}{\textbf{Model}}} & Seman. & Phono.  &Para.& Avg &  Seman. & Phono.  &Para.& Avg  & \textbf{All} \\
\hline
\multirow{4}{*}{\textbf{MMSU}}  &PAL-All-3B & 26.46&	33.58&	37.92&	33.53&	52.26	&57.83&	47.16	&53.8&	43.34 \\
 & PAL-All-3B-LAL Masked  &  25.83&	32.19&	30.1.0&	29.81	&49.82	&55.07	&47.16	&51.57	&40.34 \\
& PAL-All-3B-PLITS Masked &22.99&	33.58	&30.00	&29.57&	50.36&	54.25	&42.69	&50.87	&39.88  \\ \cline{2-11}
& PAL-3B & \textbf{29.76}&	\textbf{37.65}	&\textbf{37.52}	&\textbf{35.66}&	\textbf{58.12}&	\textbf{59.16}&	49.85&	\textbf{57.4}	& \textbf{46.18} \\ \hline

\end{tabular}
\end{table*}

\section{Preservation of Token Order Information in LAL}
\label{apx_lal_positionid}

In the standard PLITS integration paradigm, audio tokens are mapped into the LLM input space and physically prepended (or inserted) into the text token sequence. Consequently, the model assigns sequential position IDs across the entire concatenated sequence-for example, $[1, \dots, N_{sys}]$ for the system prompt, $[N_{sys}+1, \dots, N_{sys}+N_{audio}]$ for the audio tokens, and $[N_{sys}+N_{audio}+1, \dots, N_{total}]$ for the user prompt. These position IDs are used by the Rotary Positional Embeddings (RoPE) in the query (Q) and key (K) projections to encode relative and absolute positions, which is crucial for the attention mechanism to function correctly.

\begin{figure}[h]
\begin{center}
\includegraphics[width=.80\linewidth]{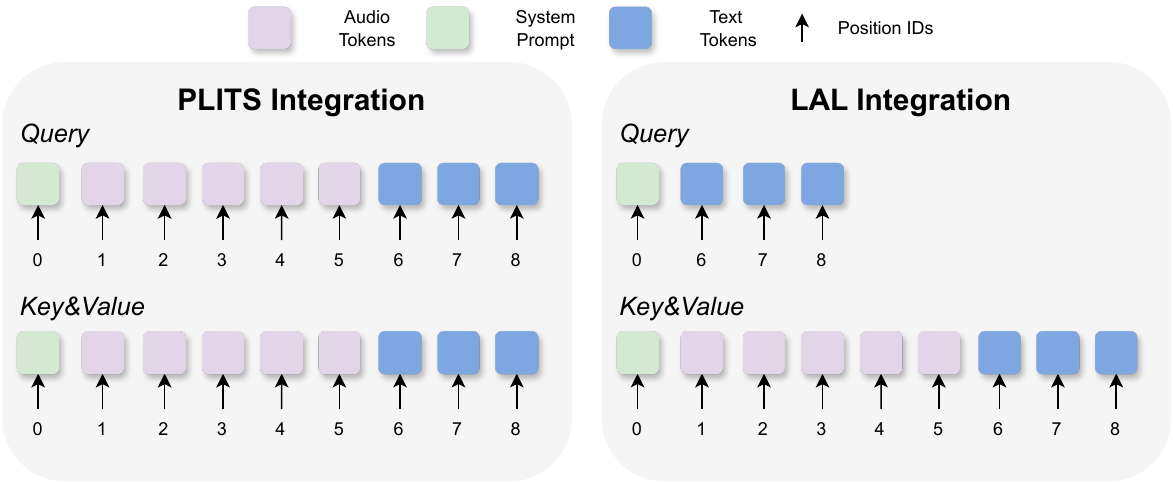}
\end{center}
\caption{LAL: Preservation of Token Order Information. This diagram illustrates how LAL preserves temporal ordering when integrating audio tokens into the LLM's attention mechanism. LAL manages position IDs by creating a gap for the audio sequence. By shifting the user prompt's position ID to $k + N_{\text{audio}} + 1$, LAL reserves the interval $[k+1, \dots, k+N_{\text{audio}}]$ for audio tokens, ensuring that text token Query, Key, and Value representations maintain identical position IDs and preserve correct self-attention structure.}

\label{fig_LAL_preserve}
\end{figure}

In our LAL implementation, audio tokens are not part of the LLM's input text sequence but are injected directly into the attention mechanism as keys and values. To ensure that the model retains accurate temporal ordering and relative distance information, we explicitly manage the position IDs to mirror the structure of PLITS.

We implement this by adjusting the position IDs of the text tokens to leave a ~\textit{gap} corresponding to the length of the audio sequence. Specifically, if the system prompt occupies indices $[1, \dots, k]$, we do not assign the immediate next integer to the user prompt. Instead, we shift the starting position ID of the user prompt to $k + N_{audio} + 1$, effectively reserving the interval $[k+1, \dots, k+N_{audio}]$ for the audio tokens. Inside the attention module, we assign these reserved position IDs to the audio keys and values as illustrated in Figure \ref{fig_LAL_preserve}.

Crucially, this adjustment ensures that for every text token, the position ID used for its Query representation is identical to the position ID used for its corresponding Key and Value representations. By maintaining this consistency, the model preserves the correct self-attention structure for text while integrating audio context at the appropriate relative positions. Similarly, we apply equivalent position ID adjustments for PAL to maintain token order integrity across all architectures.

\section{Hyper-parameters} 
\label{apx_hyp}

\subsection{Hyper-parameters: General Audio Understanding (LAL vs.\ PLITS)}
\label{apx_general_audio_hyperparam}

We evaluate PLITS and LAL integration variants under this setup. The hyper-parameters used in allo stages are summarized in Table~\ref{tab_lal_llm_training_hyper}.

\begin{table*}[htbp]
\centering
\caption{Hyper-parameters used for the three stage training of \textbf{LAL} and \textbf{PLITS} (Llama-3.2-1B)}
\small
\begin{tabular}{lccc}
\hline
\multirow{2}{*}{\textbf{Training Configuration}} & \textbf{Stage 1} & \textbf{Stage 2} & \textbf{Stage 3 $\vert$ Stage 4}\\
 & (Connector Pre training) & (LLM Fine tuning) & (LLM Fine tuning) \\
\hline
Optimizer & \multicolumn{3}{c}{AdamW~\citep{loshchilov2017decoupled}} \\
Learning Rate Schedule & \multicolumn{3}{c}{Cosine~\citep{loshchilov2016sgdr}} \\
Peak Learning Rate & 0.001 & 0.0001 & 0.0001 \\ 
Epochs & 1 & 1 & 1 \\ 
Warm up Ratio & 0.05 & 0.03 & 0.03 \\ 
Dataset Size & 1.2\,M & 1.9\,M & 5.6\,M  $\vert$ 200\,K  \\ 
Batch Size & $ 32$ & $ 12$ & $ 12$ \\ 
Gradient Accumulation Steps & \multicolumn{3}{c}{4} \\ 
GPUs & \multicolumn{3}{c}{$2 \times$ Nvidia A100 (80GB)} \\
RAM & \multicolumn{3}{c}{150\,GB} \\
Loss & \multicolumn{3}{c}{Next token loss on text part} \\
\hline
\end{tabular}
\label{tab_lal_llm_training_hyper}
\end{table*}

\subsection{Hyper-parameters: Extending to Speech and Building Unified General Audio-Music-Speech Understanding LLMs}
\label{apx_unified_hyperparam}
We evaluate PLITS, LAL, and the hybrid PAL integration variants under this setup. The hyper-parameters for both stages are summarized in Table~\ref{tab_pal_llm_training_hyper}. Despite large-scale data-parallel training, all variants optimized stably with this configuration.

\begin{table*}[htbp]
\centering
\caption{Hyperparameters used for the two stage training of \textbf{PLITS}, \textbf{LAL} and \textbf{PAL}(Llama-3.2-3B), GPU:Nvidia H100 (96GB). In case of \textbf{LAL(LT)} we trained the model for 1 additional epoch in both stages.}
\small
\begin{tabular}{lcc}
\hline
\multirow{2}{*}{\textbf{Training Configuration}} & \textbf{Stage 1} & \textbf{Stage 2} \\
 & (Connector Pre training) & (LLM Fine tuning) \\
\hline
Optimizer & \multicolumn{2}{c}{AdamW~\citep{loshchilov2017decoupled}} \\
Learning Rate Schedule & \multicolumn{2}{c}{Cosine~\citep{loshchilov2016sgdr}} \\
Peak Learning Rate & 0.001 & 0.0001 \\
Epochs & 1 & 1 \\
Warm up Ratio & 0.05 & 0.03 \\
Dataset Size & 1.7\,M & 6.4\,M \\
Batch Size & $64$ & $16$ \\
Gradient Accumulation Steps & $4$ & $6$ \\
GPUs & $32$ & $128$ \\
Loss & \multicolumn{2}{c}{Next token loss on text part} \\
\hline
\end{tabular}
\label{tab_pal_llm_training_hyper}
\end{table*}

\section{Evaluation Datasets} 
\label{apx_evaluation}

We follow the evaluation protocol of \citet{gong2024listen} for classification and captioning. Following we give details of each dataset in the evaluation suite.

\textbf{VocalSound}~\citep{gong2022vocalsound}: The VocalSound dataset consists of 21,024 crowd-sourced recordings of 6 different classes of vocal expressions collected from 3,365 unique subjects. We evaluated our model on the VocalSound evaluation set which contains 3,594 audio clips, and report top-1 accuracy scores across the 6 classes for single-class classification performance. It is important to note that VocalSound was excluded from our training data; therefore, our evaluation on VocalSound is considered zero-shot.

\textbf{ESC-50}~\citep{piczak2015esc}: The ESC-50 dataset comprises 2,000 five-second environmental audio clips categorized into 50 different classes. Following~\citet{gong2024listen}, we evaluate our model on all 2,000 audio samples and report the top-1 accuracy score for single-class classification performance. It is important to note that while ESC-50 is originally sampled from the Freesound dataset (which is included in our training data), ESC-50 itself was excluded from training. Therefore, our evaluation on this dataset is considered a weak zero-shot evaluation.

\textbf{DCASE2017 task 4} (DCASE)~\citep{Mesaros2019_TASLP}: DCASE 2017 Task 4 contains 17 sound events distributed across two categories: "Warning" and "Vehicle". The evaluation set consists of 1,350 audio clips. We evaluated our model on this dataset and report micro F1-score(MiF1) for single-class classification performance. It is important to note that DCASE 2017 task 4 is originally sampled from AudioSet, which is included in our training data. However, DCASE 2017 task 4 itself is excluded from training, making our evaluation on this dataset a weak zero-shot evaluation.

\textbf{FSD50K} (FSD) \citep{fonseca2021fsd50k}: The FSD50K evaluation set contains 10,231 audio clips. We evaluated our model on this evaluation set and report the mAP score for multi-label classification performance. Since the training and validation sets of FSD50K are included in our training data, this evaluation is considered an in-domain evaluation.

\textbf{AudioSet}~\citep{gemmeke2017audio}: We evaluated our model on this evaluation set and report the mAP score for multi-label classification performance. The training set of AudioSet is included in our training data, making this evaluation an in-domain evaluation.

\textbf{AudioCaps}~\citep{kim2019audiocaps}: The AudioCaps evaluation set contains 901 audio clips, each paired with 5 audio captions, resulting in a total of 4,505 audio-caption pairs. We evaluated our model on this evaluation set and report the captioning scores using CIDER and SPICE metrics. The training and validation sets of AudioCaps are included in our training data, making this evaluation an in-domain evaluation.

\textbf{Clotho V2}~\citep{drossos2020clotho}: The Clotho V2 evaluation set contains 1,045 audio clips, each paired with 5 audio captions, resulting in a total of 5,225 audio-caption pairs. We evaluated our model on this evaluation set and report the captioning scores using CIDER and SPICE metrics. The development and validation sets of Clotho V2 are included in our training data, making this evaluation an in-domain evaluation.

\section{LAL Integration with Frozen LLM FFN} 
\label{apx_frozen_ffn}
Standard audio-LLM training typically requires full fine tuning of the LLM. 
However, since LAL integrates audio information solely through the attention mechanism, 
we investigate whether LAL remains effective when the LLM feedforward (FFN) blocks, which are widely believed to encode much of the model’s factual and linguistic knowledge, are frozen and only the attention layers are updated. 
In Stage~2 of our training pipeline, we therefore construct a variant with the LLM FFN frozen. 
As shown in Table~\ref{tab_frozen_ffn}, performance is largely maintained under this setting. 
This result suggests that LAL can successfully integrate audio information through attention 
without modifying the knowledge stored in the FFN modules. 
Such a property has important implications for reducing training cost, improving parameter efficiency, 
and preserving the pretrained knowledge of the LLM while enabling multimodal alignment.


\begin{table*}[h]
\centering
\caption{Performance evaluation of the \textbf{LAL} Integration with frozen FFN.
Evaluation follows the protocol of \citet{gong2024listen}. AC: Audio caps, CL:Clotho AS2M: AudioSet 2M
\textsuperscript{\textdagger} indicates CIDEr and \textsuperscript{\textdaggerdbl} indicates SPICE. 
Metrics: accuracy (ESC-50, VocalSound), Mi-F1 (DCASE), and mAP (FSD, AudioSet). Complete evaluation methodology explained in Section~\ref{lal_eval_protocol} and dataset details in Appendix~\ref{apx_evaluation}}
\small
\setlength{\tabcolsep}{2.0pt}
 
\begin{tabular}{cccccccccccccc}
\hline
LLM & FFN & \multirow{2}{*}{PLITS}  & \multirow{2}{*}{LAL} & \multirow{2}{*}{LFST}  & \multicolumn{5}{c}{Classification} & \multicolumn{4}{c}{Captioning} \\
\cline{6-14}
Backbone & Frozen & & &  & ESC50 & DCASE & VS & FSD & AS2M & AC\textsuperscript{\textdagger} & CL\textsuperscript{\textdagger} & AC\textsuperscript{\textdaggerdbl} & CL\textsuperscript{\textdaggerdbl}\\
\hline

\multirow{3}{*}{Llama3.2-1B} 
& \ding{55} & \ding{51}  & \ding{55} & \ding{55}  & 64.45 & 37.69 & 51.57 & 25.23 & 9.08 & 0.59 & 0.34 & 16.30 & 10.96 \\
 & \ding{55}  & \ding{55}  & \ding{51} & \ding{55}  & ~\textbf{76.70} & ~\textbf{40.97} & ~\textbf{60.87} & ~\textbf{31.44} & ~\textbf{11.83} & ~\textbf{0.66} & 0.38 & ~\textbf{16.97} & ~\textbf{11.87} \\
& \ding{51}  & \ding{55}  & \ding{51} & \ding{55} & 71.80 & 33.99 & 55.28 & 29.38 & 10.48 & 0.63 & ~\textbf{0.40} & 16.11 & 11.75 \\
 \hline
\end{tabular}
\label{tab1}
\label{tab_frozen_ffn}
\end{table*}


\section{Extending Table~\ref{tab_lal_cls_cap} and Table~\ref{tab_lal_reasoning} with Prior audio-LLM performances}
\label{apx_cap_cls_detailed}

This section extends Table~\ref{tab_lal_cls_cap} and Table~\ref{tab_lal_reasoning} by contextualizing our results against representative prior audio-LLMs, summarized in Tables~\ref{tab_lal_prior} and~\ref{tab_lal_prior_compar}.
We stress that these cross-paper comparisons are \emph{not} strictly apples-to-apples: aside from Audio Flamingo~2, most prior systems follow the standard \textbf{PLITS} integration, and reported performance can vary substantially due to differences in training data scale, instruction-tuning mixtures, LLM and audio-encoder capacity, and evaluation protocols.
Accordingly, we report these results mainly to situate \textbf{LAL} within the broader accuracy range of existing audio-LLMs, rather than to claim a fair architectural advantage over any specific prior method.

\begin{table*}[h] 
\centering
\caption{Comparison of LAL classification and captioning performance with prior works. Except for Audio Flamingo 2, all other systems use PLITS; their higher scores mainly stem from larger datasets, bigger LLMs, and stronger audio encoders.}
\label{tab_lal_prior}

\small
\setlength{\tabcolsep}{4pt}

\begin{tabular}{cccccccccc}
\hline
\multirow{2}{*}{Models} & \multicolumn{5}{c}{Classification} & \multicolumn{4}{c}{Captioning} \\
\cline{2-10}
 & ESC50 & DCASE & VS & FSD & AS2M & AC\textsuperscript{\textdagger} & CL\textsuperscript{\textdagger} & AC\textsuperscript{\textdaggerdbl} & CL\textsuperscript{\textdaggerdbl} \\
\hline
\rowcolor{gray!10} Pengi-124M & \textbf{91.9} & 33.8 & 60.3 & 46.7 & - & - & - & - & - \\
\rowcolor{gray!10} SALMONN-7B & 16.4 & 18.0 & 16.9 & 22.1 & 13.4 & - & - & 8.3 & 7.6 \\
\rowcolor{gray!10} Audio Flamingo-2-3B & 83.9 & - & - & \textbf{47.9} & - & 0.58 & \textbf{0.46} & - & - \\
\rowcolor{gray!10} LTU-7B & 83.1 & 45.9 & 55.6 & 46.3 & 18.7 & - & - & 17 & 11.9 \\
\rowcolor{gray!10} GAMA-7B & 82.6 & 38.4 & 52.4 & 47.8 & \textbf{19.2} & - & - & 18.5 & 13.5 \\
\hline
PLITS-1B & 84.10 & 45.28 & 57.59 & 42.49 & 14.74 & 0.70 & 0.39 & 17.90 & 11.82 \\
LAL-1B (Ours) & 87.40 & 46.23 & 56.03 & 43.91 & 14.74 & 0.72 & 0.42 & 18.08 & 12.58 \\
\hline
PLITS-3B  & 84.60 & 46.16 & 59.15 & 43.29 & 15.00 & 0.70 & 0.38 & 17.90 & 12.03 \\
LAL-3B (Ours) & 89.25 & \textbf{47.21} & \textbf{60.46} & 43.86 & 15.03 & \textbf{0.73} & 0.40 & \textbf{18.61} & 12.46 \\
\hline
\end{tabular}

\end{table*}



\begin{table*}[h]
\centering
\caption{LAL performance comparison with prior works for the reasoning (CompA-R) task. All prior works use PLITS integration. Their higher scores mainly stem from larger datasets, bigger LLMs, and stronger audio encoders. }
\small
\setlength{\tabcolsep}{4pt}

 \begin{tabular}{cccccc}
 \hline
Models  & Clarity & Correctness  & Engagement & Avg \\
\hline
\rowcolor{gray!10}Qwen-Audio-Chat-8B~\citep{chu2023qwenaudioadvancinguniversalaudio} & 3.5 & 3.3 & 3.6 & 3.5 \\
\rowcolor{gray!10}LTU-7B~\citep{gong2024listen} & 3.5 & 3.2 & 3.4 & 3.4\\
\rowcolor{gray!10}SALMONN-7B~\citep{tang2024salmonn}& 2.6 & 2.4 & 2.0 & 2.3 \\
\rowcolor{gray!10}Pengi-124M~\citep{deshmukh2023pengi} & 1.8 & 1.5 & 1.3  & 1.5\\
\rowcolor{gray!10}LTU w/ CompA-R-7B~\citep{gong2024listen} & 3.5 & 3.2  & 3.4 & 3.6 \\
\rowcolor{gray!10}GAMA-IT-7B~\citep{ghosh2024gama} & 4.3 & \textbf{3.9} & \textbf{3.9} & \textbf{4.0} \\
\hline
PLITS-1B & \textbf{4.74} & \textbf{3.84}  & 2.99 & 3.86 \\
LAL-1B (Ours)  & \textbf{4.70} &3.82&3.01 & 3.80 \\ 
\hline
\end{tabular}

\label{tab_lal_prior_compar}
\end{table*}

\section{Extending Table~\ref{tab_mmau_mmar} and Table~\ref{tab_mmsu} with Prior audio-LLM performances}
\label{apx_mmau_mmar_mmsu}

This section extends the MMAU, MMAR, and MMSU evaluations by comparing our \textbf{PLITS}, \textbf{LAL}, and \textbf{PAL} unified models to representative prior audio-LLMs, as reported in Tables~\ref{tab_mmau_prior}, \ref{tab_mmar_prior}, and \ref{tab_mmsu_prior}.
We emphasize that these cross-paper comparisons are \emph{not} strictly apples-to-apples: aside from Audio Flamingo~2, most prior systems rely on the standard \textbf{PLITS} integration, and their reported performance is often shaped by substantial differences in training data scale, instruction-tuning mixtures, LLM capacity, audio encoder strength etc.
Accordingly, we include these results primarily to contextualize the absolute performance range of existing audio-LLMs on MMAU/MMAR/MMSU and to situate our methods within this broader landscape, rather than to claim a fully controlled architectural comparison against any single prior model.

\begin{table*}[h]
\centering
\caption{Evaluation on \textbf{MMAU-v05.15.25}~\citep{sakshi2024mmau} (accuracy, \%). Sound (Sn), Music (Mu), Speech (Sp). Note that values are missing (-) for LAL+ 3B as the MMAU-test online challenge was closed at the time of evaluation. Except for Audio Flamingo 2, all other systems use PLITS; their higher scores mainly stem from larger datasets, bigger LLMs.  $^{\dagger}$ \textit{LT indicates the same model trained for more steps (no architectural changes); excluded from bolding to ensure a fair comparison.}
}
\small
\setlength{\tabcolsep}{4pt}
\begin{tabular}{lcccccccc}
\hline
\multicolumn{1}{c}{} & \multicolumn{2}{c}{Sn} & \multicolumn{2}{c}{Mu} & \multicolumn{2}{c}{Sp} & \multicolumn{2}{c}{Total (Avg)} \\
\multicolumn{1}{c}{\multirow{-2}{*}{Model}} & mini & test & mini & test & mini & test & mini & test \\
\hline
\rowcolor{gray!10} Step-Audio-2-mini-8.3B~\citep{wu2025stepaudio2technicalreport} & 79.30 & 75.57 & 68.44 & 66.85 & 66.18 & 66.49 & 72.73 & 70.23 \\
\rowcolor{gray!10} DeSTA2.5-Audio-8B~\citep{lu2025desta25audiogeneralpurposelargeaudio} & 70.27 & 66.83 & 56.29 & 57.10 & \textbf{71.47} & \textbf{71.94} & 66.00 & 65.21 \\
\rowcolor{gray!10} SALMONN-13B~\citep{tang2024salmonn}  & 41.14 & 42.10 & 37.13 & 37.83 & 26.43 & 28.77 & 34.90 & 36.23 \\
\rowcolor{gray!10} GAMA-7B~\citep{ghosh2024gama} & 31.83 & 30.73 & 17.71 & 17.33 & 12.91 & 16.97 & 20.82 & 21.68 \\
\rowcolor{gray!10} GAMA-IT-7B~\citep{ghosh2024gama} & 30.93 & 32.73 & 26.74 & 22.37 & 10.81 & 11.57 & 22.83 & 22.22 \\
\rowcolor{gray!10} LTU-7B~\citep{gong2024listen} & 20.42 & 20.67 & 15.97 & 15.68 & 15.92 & 15.33 & 17.44 & 17.23 \\

\rowcolor{gray!10} Qwen2.5-Omni-7B~\citep{xu2025qwen25omnitechnicalreport} & 78.10 & \textbf{76.77} & 65.90 & 67.33 & 70.60 & 68.90 & 71.50 & 71.00 \\
\rowcolor{gray!10} Qwen2-Audio-Instruct-7B~\citep{chu2024qwen2audiotechnicalreport} & 67.27 & 61.17 & 56.29 & 56.29 & 55.67 & 55.57 & 59.90 & 57.40 \\
\rowcolor{gray!10} M2UGen-7B~\citep{liu2024m2ugenmultimodalmusicunderstanding} & 43.24 & 42.44 & 37.13 & 38.53 & 35.37 & 35.77 & 37.90 & 39.76 \\
\rowcolor{gray!10} MusiLingo-7B~\citep{deng2024musilingobridgingmusictext} & 43.24 & 41.93 & 40.12 & 41.23 & 31.23 & 31.73 & 38.10 & 38.29 \\

\rowcolor{gray!10} Audio Flamingo-3-8.2B~\citep{goel2025audio}&  \textbf{79.58} & 75.83 & \textbf{73.95} &\textbf{ 74.47} & 66.37 & 66.97 & 73.30 & 72.42 \\
\rowcolor{gray!10} Audio Flamingo-2-3B~\citep{ghosh2025audio} & 71.47 & 68.13 & 70.96 & 70.20 & 44.74 & 44.87 & 62.40 & 61.06 \\
\rowcolor{gray!10} Audio Flamingo Chat-1B~\citep{kong2024audio}& 25.3 & 23.33 & 17.66 & 15.77 & 6.91 & 7.67 & 16.60 & 15.59 \\

\hline

 PLITS-3B & 75.68 & 72.03 & 70.96 & 69.63 & 46.25 & 46.48 & 64.30 & 62.91 \\
 LAL-3B & 73.57 & 69.53 & 67.66 & 69.67 & 47.15 & 50.07 & 62.80 & 63.24 \\
 \cellcolor{gray!15}LAL-3B (LT)$^{\dagger}$ & \cellcolor{gray!15}75.08 & \cellcolor{gray!15} -  & \cellcolor{gray!15}70.66 & \cellcolor{gray!15} - & \cellcolor{gray!15}48.05 & \cellcolor{gray!15} - & \cellcolor{gray!15}64.60 & \cellcolor{gray!15} - \\
 PAL-3B & 78.08 & 74.60 & 73.05 & 71.30 & 59.46 & 58.36 & 70.20 & 68.20 \\ \hline

\end{tabular}
\label{tab_mmau_prior}
\end{table*}


\begin{table*}[h]
\centering
\caption{Evaluation of PAL on ~\textbf{MMAR}~\citep{ma2025mmar}  (accuracy, \%). Abbr:  Sound (Sn), Music (Mu), Speech (Sp). Except for Audio Flamingo 2, all other systems use PLITS; their higher scores mainly stem from larger datasets, bigger LLMs.  $^{\dagger}$ \textit{LT indicates the same model trained for more steps (no architectural changes); excluded from bolding to ensure a fair comparison.}}
\small
\setlength{\tabcolsep}{4pt}

\begin{tabular}{lcccccccc}
\hline
\multicolumn{1}{c}{\multirow{2}{*}{Models}} & \multicolumn{1}{c}{\multirow{2}{*}{Sn}} & \multicolumn{1}{c}{\multirow{2}{*}{Mu}} & \multicolumn{1}{c}{\multirow{2}{*}{Sp}} & \multicolumn{1}{c}{Mix} & \multicolumn{1}{c}{Mix} & \multicolumn{1}{c}{Mix} & \multicolumn{1}{c}{Mix} & \multicolumn{1}{c}{Total} \\
\multicolumn{1}{c}{} & \multicolumn{1}{c}{} & \multicolumn{1}{c}{} & \multicolumn{1}{c}{} & \multicolumn{1}{c}{Sn-Mu} & \multicolumn{1}{c}{Sd-Sp} & \multicolumn{1}{c}{Mu-Sp} & \multicolumn{1}{c}{Sn-Mu-Sp} & \multicolumn{1}{c}{Accuracy} \\ \hline
\rowcolor{gray!10} Audio Flamingo-2-3B & 24.85	&17.48	&20.75	&18.18	&26.61	&23.17	&8.33	&21.90 \\
\rowcolor{gray!10} Audio Flamingo-3-8.2B & -  & -  & -  & -  & -  & -  & - & \textbf{58.5}\\
\rowcolor{gray!10} LTU-7B & 19.39	&19.90	&13.95	&18.18	&24.77	&21.95	&16.67	&19.20 \\
\rowcolor{gray!10} SALMONN-13B & 30.30	&31.07	&34.69	&9.09	&34.86	&35.37	&41.67	&33.20 \\
\rowcolor{gray!10} GAMA-7B & 29.09	&24.27	&27.89	&27.27	&24.77	&28.05	&20.83	&26.50 \\
\rowcolor{gray!10} GAMA-IT-7B & 22.42	 &16.02	 &12.24	&36.36	&22.48	&14.63	&12.50	&17.40 \\
\rowcolor{gray!10} Qwen2.5-Omni-7B &\textbf{58.79}	& \textbf{40.78	}& \textbf{59.86}	& \textbf{54.55}	& \textbf{61.93}	& \textbf{67.07} 	& \textbf{58.33}	&56.70 \\
\hline

PLITS-3B & 38.79 & 40.29 & 37.41 & {36.36 }& 48.17 & 40.24 & {50.00 }& 41.10  \\
 LAL-3B & 38.18 & {42.72} & 35.37 & 18.18& 41.28 & 39.02 & 41.67 &38.90 \\
 \cellcolor{gray!15}LAL-3B (LT)$^{\dagger}$  & {\cellcolor{gray!15}47.88} & {\cellcolor{gray!15}43.69} & {\cellcolor{gray!15}38.10} & \cellcolor{gray!15} 27.27&  \cellcolor{gray!15}46.44& \cellcolor{gray!15} 45.12& \cellcolor{gray!15}54.17 &{\cellcolor{gray!15}43.50} \\
 PAL-3B & 44.85 & 41.75 &  43.54  &27.27& 51.83 &  47.78 &41.67& {45.40} \\
 \hline
\end{tabular}

\label{tab_mmar_prior}
\end{table*}
 

\begin{table*}[h]
\centering
\caption{Evaluation on \textbf{MMSU}~\citep{wang2025mmsu} across perception and reasoning
dimensions in Semantics (Seman.), Phonology (Phono.), and Paralinguistics (Para.) domains. Except for Audio Flamingo~2, all other systems adopt PLITS; most are trained on substantially larger, speech-heavy datasets than ours and use larger LLMs. $^{\dagger}$ \textit{LT indicates the same model trained for more steps (no architectural changes); excluded from bolding to ensure a fair comparison.}
}
\small
\setlength{\tabcolsep}{4pt}
\begin{tabular}{lccccccccc}
\hline
\multicolumn{1}{c}{} & \multicolumn{4}{c}{Perception} & \multicolumn{4}{c}{Reasoning} & Avg \\
\multicolumn{1}{c}{\multirow{-2}{*}{Model}} & Seman. & Phono.  &Para.& Avg &  Seman. & Phono.  &Para.& Avg  & All \\
\hline

\rowcolor{gray!10} LTU-7B~\citep{gong2024listen}  & 21.34  & 22.46  & 18.73  &20.81  &22.65& 25.53 & 24.74 &24.37  & 22.61 \\
\rowcolor{gray!10} LTU-AS~\citep{gong2023joint}  & 25.89  & 24.71 & 21.64 & 24.13 & 26.53 &  25.68 & 25.04 & 25.92  & 25.03 \\ 
\rowcolor{gray!10} SALMONN-13B~\citep{tang2024salmonn}   &31.55 & 29.08 & 28.71 & 29.83 & 36.43 & 26.22  & 25.26 & 30.04 &   30.01 \\
\rowcolor{gray!10} Qwen2-Audio-Instruct-7B~\citep{chu2024qwen2audiotechnicalreport}  & 52.14&32.87&35.56 &39.02&77.62&64.81&46.67 &68.90 & 53.27 \\
\rowcolor{gray!10} Qwen2.5-Omni-3B~\citep{xu2025qwen25omnitechnicalreport}  & 52.04  &\textbf{38.73} & 39.19&42.37&81.20&81.12 &41.19&72.76 & 56.83\\
\rowcolor{gray!10} Qwen2.5-Omni-7B~\citep{xu2025qwen25omnitechnicalreport}  &\textbf{55.12} &37.33 &\textbf{39.35} &\textbf{42.50}&\textbf{88.00} &\textbf{81.37} &48.36 &\textbf{79.83} & \textbf{60.57}\\
\rowcolor{gray!10} Audio Flamingo-3-8.2B ~\citep{goel2025audio} & - & -  &-  & - & - &- & - &- & 61.4\\

\hline

 PLITS-3B &24.09 &	31.98 &	34.75 &	31.12 &	49.46	 &52.3	 &42.99	 &49.71 &	40.12 \\
 LAL-3B  & 26.61	&33.8	&31.19	&31.01&	50.00	&54.35	& \textbf{52.54}	&52.11&	41.22  \\
\rowcolor{gray!15} LAL-3B (LT)$^{\dagger}$ & 31.34 &34.33 & 27.13 & 30.78 & 53.97 & 58.75 & 51.34 & 55.54 & 42.76 \\
 PAL-3B & 29.76&	37.65 &37.52	& {35.66}&	 58.12 &	59.16 &	49.85&	 57.4	& 46.18 \\ \hline

\end{tabular}
\label{tab_mmsu_prior}
\end{table*}

\section{Extended literature review}
\label{apx_ext_lit_review}
\paragraph{Audio-LLM Datasets} Beyond architecture, recent works have focused on  high-quality instruction tuning datasets, both open-source and proprietary~\citep{goel2025audio,ghosh2024gama,chu2024qwen2audiotechnicalreport,xu2025qwen25omnitechnicalreport} and build audio reasoning benchmarks~\citep{sakshi2024mmau,deshmukh2025audio,deshmukh2025adiff}. Training PLITS or Flamingo-style models on these resources improves instruction following and audio reasoning, with most gains driven by the data rather than the integration scheme.

\paragraph{Audio Representation learning} To obtain rich semantic audio representations, recent advances in audio representation learning have led to powerful audio encoders trained with diverse objectives across different pretraining paradigms. The studies have shifted from simple supervised learning paradigms ~\citep{gong2022contrastive,gong2021ast} to more complex self-supervised paradigms ~\citep{huang2022amae, Ahmed_2024, chen2024eat, alex2025sslam} that employ contrastive objectives and masked-token prediction strategies to capture both global semantic structure and fine-grained local details within audio representations. Furthermore, in the multimodal pretraining paradigm, language-aligned audio representations are obtained through contrastive audio–language models~\citep{elizalde2023clap,wu2023large,ghosh2025reclap} which align the representations of audio and language into a unified semantic space. Transcription-based approaches ~\citep{radford2023robust} leverage next token prediction on speech-to-text tasks to learn robust audio representations that capture speech semantics and acoustic-linguistic relationships.

 \section{LFST Connector: Language aligned and Fine grained Spatiotemporal Connector}
\label{apx_lfst}

We adopt the connector \emph{proposed in Cambrian}~\citep{tong2024cambrian} and apply it in our audio setting to fuse a language aligned encoder such as CLAP~\cite{elizalde2023clap} with a self supervised encoder such as SSLAM~\cite{alex2025sslam}. The connector produces a compact set of latent tokens that combine semantic cues from CLAP with fine grained spatiotemporal detail from SSLAM, while keeping sequence length fixed and avoiding the overhead of naive concatenation.

\begin{figure}[h]
\begin{center}
\includegraphics[width=0.80\linewidth]{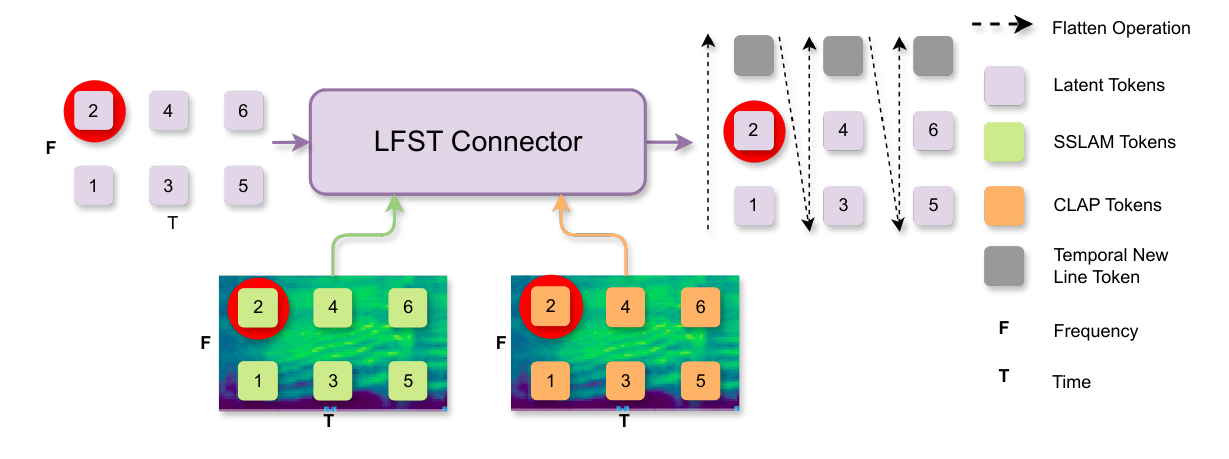}
\end{center}
\caption{Overview of \textsc{LFST} using the Cambrian connector~\citep{tong2024cambrian}. A single latent token is broadcast to every time–frequency location and then updated inside the connector by cross attention with local \textsc{SSLAM} and \textsc{CLAP} features, fusing fine grained spatiotemporal detail with language aligned semantics. The red tokens illustrate the latent query and the local encoder keys and values it attends to. A newline token is inserted at each new time step so the flattened sequence preserves the original spatiotemporal layout while keeping the output length fixed.}
\label{fig_lfst_overview}
\end{figure}

\paragraph{Formalization.}
Let the encoder outputs be
\[
H_{\text{sslam}}, \; H_{\text{clap}} \in \mathbb{R}^{F \times T \times d}, \quad z \in \mathbb{R}^{d},
\]
where $F$ is frequency, $T$ is time, and $d$ is the feature dimension. Following \citet{tong2024cambrian}, a single latent token $z$ is broadcast to each spatiotemporal location, yielding $z_{f,t}$ for every $(f,t)$. Inside the connector, which consists of 3 cross attention layers, each $z_{f,t}$ is updated through cross attention with the corresponding local regions of $H_{\text{sslam}}$ and $H_{\text{clap}}$. To preserve temporal structure when flattening across $(F,T)$, we insert a \emph{newline token} along the frequency axis so that each new time step begins with this marker before its spectral tokens (see Figure~\ref{fig_lfst_overview}).

\end{document}